\newif\iftwocol
\twocolfalse

\iftwocol
  \documentclass[pdflatex,sn-apa,iicol]{sn-jnl}
\else
  \documentclass[pdflatex,sn-apa]{sn-jnl}
\fi

\usepackage{graphicx}
\usepackage{multirow}
\usepackage{amsmath,amssymb,amsfonts}
\usepackage{amsthm}
\usepackage{mathrsfs}
\usepackage[title]{appendix}
\usepackage{xcolor}
\usepackage{textcomp}
\usepackage{manyfoot}
\usepackage{booktabs}
\usepackage{tabularx}
\usepackage{ragged2e}
\usepackage{placeins}

\iftwocol
  \makeatletter
  
  \renewenvironment{figure}[1][]{\@dblfloat{figure}[!tbp]}{\end@dblfloat}
  \makeatother
  \newcommand{\colurl}[1]{{\urlstyle{same}\url{#1}}}
  \AtBeginDocument{}
\fi

\newcommand{\AdjAgreeSplit}{0.56 $[0.51, 0.61]$ (192/341)}
\newcommand{\AdjAgreeAbstain}{0.77 $[0.70, 0.82]$ (151/197)}
\newcommand{\AdjAgreeUnanimous}{0.79 $[0.69, 0.86]$ (63/80)}
\newcommand{\AdjKappaSplit}{0.01}
\newcommand{\AdjHumanBreach}{494}
\newcommand{\AdjGraderBreach}{402}
\newcommand{\AdjPrecision}{0.85 $[0.81, 0.88]$ (342/402)}
\newcommand{\AdjRecall}{0.69 $[0.65, 0.73]$ (342/494)}
\newcommand{\AdjRetestAgree}{0.93 $[0.87, 0.96]$ (115/124)}
\newcommand{\AdjRetestKappa}{0.80}
\newcommand{\AdjSensitivity}{Replacing the grader's label with the annotator's on the 584 of 2{,}400 genetic-search rows whose response was adjudicated changes 153 labels to breach and 60 to non-breach. Attack success rises by between 0.00 and 0.06 per safeguard, $\phi_{\mathrm{null}}$ moves from 0.620 to 0.594, and no non-guard pairs lie above it, the largest excess being $-0.017$ against $+0.026$ on the grader's labels.}

\begin{document}

\title[Diversity of Layered Safeguards]{How Diverse Are Layered Safeguards?
Estimands and Measurement for Ensembles of LLM Jailbreak Defenses}

\author*[1,2]{\fnm{Abrar} \sur{Alotaibi}}\email{amotaibi@iau.edu.sa}
\author[1,3]{\fnm{Moataz} \sur{Ahmed}}

\affil[1]{\orgdiv{Information and Computer Science Department},
\orgname{King Fahd University of Petroleum and Minerals},
\orgaddress{\city{Dhahran}, \country{Saudi Arabia}}}
\affil[2]{\orgdiv{Department of Computer Science, College of Computer Science
and Information Technology}, \orgname{Imam Abdulrahman Bin Faisal University},
\orgaddress{\city{Dammam}, \country{Saudi Arabia}}}
\affil[3]{\orgdiv{SDAIA-KFUPM Joint Research Center for Artificial
Intelligence}, \orgname{King Fahd University of Petroleum and Minerals},
\orgaddress{\city{Dhahran}, \country{Saudi Arabia}}}

\abstract{Practitioners protect large language models (LLMs) by stacking safeguards: a likelihood
filter, a guard classifier, hidden-state probes, a decoding-time intervention.
Such a stack is an ensemble of learned detectors, and ensemble theory states
when combination pays: when members fail on different inputs. Stacking is
recommended throughout the LLM security literature, but this condition has not
been measured for any pair of safeguards.

The condition is three quantities, not one, and conflating them is why it has
resisted measurement. Separately optimized attacks license only a
claim about which \emph{behaviors} are jointly vulnerable; the identity that
governs a deployed stack is written over shared \emph{inputs}; and neither
predicts what an adversary attacking the assembled pipeline achieves. We define
the three estimands, give an estimator for each, and supply two controls: the
correlation between two independent runs of the same attack, which
sets the level a pair must clear, and a multilevel model separating detector
diversity from behavior difficulty.

On a seven-layer stack and two open-weight targets, no pair of
safeguards is distinguishable from the correlation that two independent runs of
the attack produce with no safeguard present, and behavior difficulty explains much
of what remains. A second attack family, an attacker
model rewriting its own prompts, reaches that verdict independently on a
thirtieth of the budget and overturns the practical one: the only member that
blocked the first adversary admits 0.19 under the second, against 0.03.
Diversity between neural safeguards is measurable, but it and any member's
strength are properties of a (stack, adversary, target) triple.}


\maketitle

\section{Introduction}\label{sec:intro}

A deployed language model is rarely defended by one mechanism. Production systems
place several detectors around the model: a filter on the input, a guard model
that classifies prompts or responses, sometimes a probe reading the model's own
activations, sometimes an intervention at decoding time. The reasoning is rarely
stated: if one detector misses, another may catch.

That reasoning is an ensemble argument, and the ensemble literature established its
governing condition. Combining classifiers pays to the extent that
the members fail on \emph{different} inputs, the condition is measurable through
pairwise and non-pairwise diversity statistics, and it must be verified, not
inferred from the fact that two members look different
\citep{kuncheva2003measures, brown2005diversity, tang2006analysis}. The
dependability literature reached the same conclusion for redundant components
under a common threat \citep{littlewood2001modeling,
littlewood2004redundancy}. Layering is standard advice in the security
literature on large language models (LLMs), which has not measured the condition for any
pair of safeguards.

\paragraph{The measurement problem.}
The obvious experiment is to attack each defense, record which behaviors each
one lets through, and correlate the resulting vectors. That experiment answers a
different question from the one it appears to answer.

An adaptive adversary re-optimizes against each defense. The prompt that defeats
a likelihood filter is not the prompt that defeats a guard model, so the two
breach vectors are indexed by \emph{behavior}, not by input. The identity that
governs a deployed stack is written over a shared input: both
members see the same prompt, and the stack admits it only if both admit it.
The two quantities coincide only when attacks transfer perfectly between
defenses. A third quantity, what an
adversary achieves against the assembled pipeline, is implied by neither,
because the layers alter the feasible attack region and, in sequence, the input
that later layers observe.

\paragraph{Contributions.}
This paper is a measurement contribution and proposes no new defense.

\begin{itemize}
\item \textbf{Three estimands.} We define behavior-level
co-vulnerability, prompt-level co-failure and assembled-stack residual, give the
conditions under which each is the right quantity, and show that the first does
not bound the others (Section~\ref{sec:ensemble}).

\item \textbf{An estimator with its own null.} We
estimate the correlation between two independent runs of the same search against
the \emph{undefended} model, use it as the level that any defense pair must
exceed, and report diversity corrected for that unreliability
(Section~\ref{sec:controls}).

\item \textbf{A model that separates diversity from difficulty.} Behaviors differ
in how readily they are elicited, which correlates any two
detectors regardless of mechanism. We fit a random-intercept model over defenses,
behaviors and replicates, and report the association that difficulty alone
predicts alongside the excess over it (Section~\ref{sec:controls}).

\item \textbf{Measurements on a seven-layer stack over two open-weight targets},
with three independent attack replicates each, two attack families, a held-out
over-refusal set, measured grader reliability, grader abstention separated from grader
judgment, and targeted human adjudication (Section~\ref{sec:results}).

\item \textbf{Evidence on what hidden-state probes detect.} A linear probe on
mid-layer activations separates harmful from benign prompts at an area under the receiver operating
characteristic curve (AUROC) of 0.93 and
blocks none of the attacked prompts, because the attack moves its score
into the range it assigns to benign prompts. An early-layer probe separates two
harmless corpora better than it separates harm (Section~\ref{sec:probes}).
\end{itemize}

\paragraph{Summary of findings.}
The main result is negative. Two independent runs of the same attack against the
undefended model produce breach vectors that agree at $\phi_{\mathrm{null}} = 0.620$,
and no pair of safeguards exceeds that level by an amount the measurement can
resolve: five of the twenty-one lie above it, by at most $0.026$, against a null
whose own three estimates span $0.094$ and pair intervals with a median width of
$0.343$. Conditioning on how hard each behavior is to elicit from the undefended
model removes 71\% of the association among the non-guard pairs. The members are not
independent, but this design cannot attribute their shared structure to the pairing
of mechanisms.

The other two estimands and the two replications qualify that result. On
shared inputs, four of the seven members return the target's own response to almost
every adversarial prompt (Section~\ref{sec:e2}). The estimate composed from members
measured one at a time is below the assembled stack's measured residual in every
comparison that has resolution, and it understates the false-refusal rate, $0.60$
against a measured $0.83$ (Sections~\ref{sec:e3} and~\ref{sec:frr}). Under the genetic
attack the guard classifier reduces attack success from $0.74$ to $0.03$, the
perturbation aggregator lowers it to $0.65$, and the other five safeguards leave it
inside the undefended estimate's interval; under a second attack family on one
thirtieth of the search budget the guard classifier admits $0.19$
(Section~\ref{sec:family}). On a more strongly aligned target the attack succeeds too
rarely for the null to be estimated precisely, the share of association attributable
to behavior difficulty falls by about half, and on identical inputs the perturbation
aggregator no longer lowers attack success (Section~\ref{sec:replication}).

The paper proceeds as follows. Section~\ref{sec:related} places the question in the
ensemble, dependability and LLM-safeguard literatures. Section~\ref{sec:ensemble} defines
the three estimands, and Section~\ref{sec:design} describes the experiment and the controls
the estimates need. Section~\ref{sec:results} reports the measurements,
Section~\ref{sec:discussion} discusses their use and their limits, and
Section~\ref{sec:conclusion} concludes.

\section{Related work}\label{sec:related}

Three bodies of work bear on the question: the literature on diversity in classifier
ensembles and redundant systems, which supplies the condition and the statistics; the work
on safeguards for language models, which combines them without measuring the condition; and
adaptive evaluation, which determines how the measurement has to be made.

\subsection{Diversity in ensembles of classifiers}

That combining classifiers pays only to the extent that their errors differ is one of the
older results in ensemble learning, and the literature that established it also supplied the
instruments. Pairwise statistics (the $Q$-statistic, the disagreement measure, the
double-fault rate and the correlation of error indicators) and non-pairwise ones (entropy,
the Kohavi--Wolpert variance, interrater $\kappa$ and a difficulty measure) were catalogued
and compared by \citet{kuncheva2003measures}, who also showed that diversity tracks ensemble
accuracy only imperfectly, so a diversity figure is evidence about an ensemble and not a
prediction of it. A subfield followed on \emph{creating} diversity
\citep{brown2005diversity, sagi2018ensemble}, because independent members
are hard to obtain.

The dependability literature reached the same conclusion for redundant components facing a
common threat. \citet{littlewood2001modeling} model design diversity and show that the
reliability of a redundant system is governed by coincident failure, not by the
reliability of its parts, and \citet{littlewood2004redundancy} carry the argument into
security, where the common cause is an adversary. That is the
reference class for a safeguard stack: the members are deployed together, against an opponent
who is free to look for what they share. The adversarial branch of classifier fusion makes the
same point constructively, since an attacker who targets a shared property defeats a multiple
classifier system at the cost of defeating one member \citep{biggio2010multiple,
biggio2014security}.

This paper asks whether those results hold for neural safeguards around a language model.
It faces a question those literatures did not: when the adversary re-optimizes
against each member separately, the error indicators are no longer defined on a common input,
and the diversity statistic changes meaning.

\subsection{Safeguards for language models, and their combination}

Individual safeguards are well studied. Surface statistics of the prompt
\citep{jain2023baseline}, a separate classifier that judges prompts or responses against a
policy \citep{inan2023llama}, interventions on the decoding distribution
\citep{xu2024safedecoding}, and aggregation over perturbed copies of the input
\citep{robey2023smoothllm} are all standard, and each is normally evaluated alone.

A smaller body of work combines them. \citet{lu2024autojailbreak} propose a
mixture-of-defenders in which experts specialize by attack class; guardrail ensembles combine
several classifiers by boosting, by a mixture of experts, or by logical reasoning over their
outputs \citep{dualbreach2025, aegis2024, r2guard2025}; and the operational literature simply
recommends stacking, on the explicit grounds that the mistakes of different defenses are
likely to be uncorrelated \citep{operationalizingthreat2025}. Each of these assumes what this
paper measures: the premise that the combined components fail on different inputs is argued
from the mechanisms' apparent dissimilarity rather than estimated.

Recent work has begun to attack pipelines rather than components. \citet{mckenzie2025stack}
stage attacks against a safeguard pipeline and defeat it by breaking its stages in sequence;
\citet{luo2026cascade} evaluate many attack and defense combinations systematically; and
\citet{wang2026sok} systematize the evaluation of jailbreak guardrails and document how
inconsistent the field's protocols are. These establish that pipelines fail and that
evaluation is uneven. None of them estimates diversity between members, and none separates
the estimands of Section~\ref{sec:estimands}.

\subsection{Adaptive evaluation}

Defenses that appear robust against a fixed attack set have been broken repeatedly once the
attacker optimizes against the deployed configuration, in adversarial machine learning
generally and for language models specifically \citep{andriushchenko2024jailbreaking,
attackermovessecond2025}. We take this as a design constraint: every
number reported here comes from an adversary with knowledge of the defense it faces, and the
one defense that cannot be made adaptive within our budget is reported as such.

Each member we test was published with an
evaluation against the attacks available when it was proposed, for several of them
an unconstrained gradient suffix search. We re-evaluate them against a fluent search
that optimizes the quantity each one computes. This is a harder test than any of them
was originally set, and the results are what that test yields, not a correction of the
original papers.

\section{A safeguard stack as an ensemble of learned detectors}\label{sec:ensemble}

This section treats the stack as an ensemble. It describes the members and what they
share, states the identity that decides when combining them pays, and separates that
identity into the three quantities an experiment can estimate.

\subsection{The members are learned, and they share a substrate}

The stack we study has seven members, and the distinction that matters for what follows is
whether a member is itself a learned model. We write \emph{safeguard}, \emph{defense} and
\emph{member} interchangeably for one component of the stack, and \emph{layer} where the order
in which they run is what matters.

Three are. A guard model is an 8B network fine-tuned to classify prompts and responses
against a safety policy \citep{inan2023llama}. Two linear probes read the target's own hidden
states, one at a middle layer and one early, and classify from the representation directly.
A fourth member, a windowed likelihood filter, is learned only in the sense that it thresholds
the target network's own token likelihoods \citep{jain2023baseline}; we treat it with the
learned members because what it measures is a property of the network.

Three are not learned, and they are the controls the comparison needs. A refusal-priming
system prompt acts on the target's first-token output distribution without any trained
component, and a perturbation aggregator \citep{robey2023smoothllm} queries the network
several times on character-perturbed copies and votes. The third, a surface token-anomaly
statistic, involves no model at all and serves as the non-neural contrast: if diversity were
simply a matter of
mechanisms looking different, this member should be the most diverse of all.

This is not the textbook ensemble. There, members are estimated from overlapping draws
of a common training distribution, their errors covary because of that shared sample, and the
remedies all widen the pool: resample the data, perturb the representation, vary the
hypothesis class, penalize agreement during training \citep{brown2005diversity,
sagi2018ensemble}. Here several members are not estimated at all, and apart from the two
probes those that are share no training data with each other. What they share is the
\emph{object they operate on}. Every one
of them wraps the same target network, and safety behavior in such a network is concentrated
in its output distribution over the first few generated tokens \citep{qi2024safety}. An input
that moves the model off that behavior is simultaneously an input on which a likelihood
filter, a probe trained on representations of such inputs, and a smoothing wrapper relying on
their instability are each more likely to be wrong.

Whether that shared substrate shows up as correlated failure is an empirical
question. Behavior difficulty,
benchmark construction, the attack objective and the shared grader are competing
explanations for any correlation we observe, and Section~\ref{sec:controls} describes how we
separate the first of them.

\subsection{When combination pays}

The quantity a deployer wants is the residual attack success rate (ASR) of the stack. Write
$b_\ell(i) \in \{0,1\}$ for the event that layer $\ell$ is breached on input $i$,
and $p_\ell = \Pr[b_\ell = 1]$. For two layers evaluated on the same input, the
residual is the probability that both are breached, which for binary variables is
exactly
\begin{equation}\label{eq:joint}
\mathrm{ASR}_{\ell_1 \ell_2}
\;=\; p_1 p_2 \;+\; \rho\,\sqrt{p_1(1-p_1)\,p_2(1-p_2)},
\end{equation}
where $\rho$ is the correlation of the two breach indicators, equal to their
Matthews coefficient $\phi$ and to one of the standard pairwise diversity
measures of classifier fusion \citep{kuncheva2003measures}. The multiplicative
assumption that motivates stacking is the special case $\rho = 0$, and the excess
$\Delta = \mathrm{ASR}_{\ell_1\ell_2} - p_1p_2$ measures how far a stack falls
short of it.

A positive $\rho$ does not make stacking pointless. Adding a second member to a first reduces
residual attack success by $p_1\,(1 - \Pr[b_2 = 1 \mid b_1 = 1])$, which is
positive whenever the second member blocks anything that the first admits.
Correlation shrinks that gain without removing it, so the
question is how much a stack
buys relative to the independent case that motivated it.

Two properties of Eq.~\eqref{eq:joint} limit its use. First, it is an
identity for two indicators evaluated \emph{in parallel on the same input}; a
deployed stack runs its members in sequence, and a member that perturbs the
prompt or prepends a prefix changes what the next member sees. Second, $\phi$ on
binary indicators cannot reach $1$ unless the marginals agree, so it is not
comparable across pairs with different $p_1, p_2$ without normalization.

\subsection{Three estimands}\label{sec:estimands}

The distinction this paper turns on is which inputs the two indicators are
evaluated on. Table~\ref{tab:estimands} states the three quantities, and every
claim in Section~\ref{sec:results} is labeled with the one it rests on.

\begin{table}[!htbp]
\centering
\caption{The three estimands and the question each answers.}
\label{tab:estimands}
\small
\begin{tabularx}{\textwidth}{@{}lXX@{}}
\toprule
Estimand & What is held fixed & Question it answers \\
\midrule
E1, behavior-level co-vulnerability
& the behavior; each defense faces a prompt optimized against itself
& Are the same harmful behaviors reachable past each defense, when the adversary
is free to retarget? \\
\addlinespace
E2, prompt-level co-failure
& the input; both defenses see the same prompt
& The quantity Eq.~\eqref{eq:joint} is written over: do two defenses fail
together on one input? \\
\addlinespace
E3, assembled-stack residual
& nothing; the adversary attacks the deployed pipeline
& What does a deployer actually face? \\
\bottomrule
\end{tabularx}
\end{table}

E1 does not bound E2. The two coincide only if an attack optimized against one defense
transfers perfectly to another, and attacks need not: a prompt shaped to pass a likelihood
filter is fluent by construction, which does nothing to help it past a semantic classifier.
E1 can therefore be large while E2 is small, if each defense is separately defeated but by
inputs that the other would catch, and it can be small while E2 is large.

Neither bounds E3. Because the members run in sequence, the set of inputs that
satisfy every member at once can be far smaller than the intersection of the sets that satisfy
each alone, so intersecting independently measured breach
vectors does not predict the assembled stack in either direction.

The distinction is easily lost in wording. An earlier version of
this work, like the literature it drew on, estimated E1 while describing defenses that
``fail on the same inputs'', when the inputs differed by
construction. We therefore report all three estimands and do not treat one as a proxy
for another.

\section{Design}\label{sec:design}

The experiment estimates the three quantities of Section~\ref{sec:estimands} on one stack.
We describe the targets, behaviors and safeguards, then the adversaries, then the controls
without which the estimates cannot be read, and finally how the assembled stack is defined
and attacked.

\subsection{Targets, behaviors and safeguards}

The primary target is Vicuna-7B-v1.5 \citep{zheng2023judging}, in half precision, with a
replication on the more strongly aligned Llama-2-7b-chat \citep{touvron2023llama2}. The weakly
aligned target is primary deliberately: a defense can only be assessed for correlated failure
if it fails sometimes, and against a refusal-robust target several members have breach rates
near zero, at which point the diversity statistics are undefined, not small
(Section~\ref{sec:replication}).

Behaviors are the 100 harmful prompts of JailbreakBench \citep{chao2024jailbreakbench} with
its matched benign set, and false refusals are additionally measured on a held-out sample of
400 prompts drawn from OR-Bench \citep{cui2024or}, so that the benign side is not confined to
prompts built as counterparts to the harmful ones.

Seven safeguards are deployed (Table~\ref{tab:defenses}). Two occupy each of two
mechanism families, so that within-family pairs exist to compare against across-family ones.
Every thresholded member is calibrated to block 1\% of an \emph{external} corpus of 805
general instructions, disjoint from both the evaluation set and the probes' training pool.
Section~\ref{sec:probes} shows that even this operating point does not transfer across
benign corpora.

\begin{table}[!htbp]
\centering
\caption{The deployed safeguards, with the short names used in later tables and figures. ``Learned'' marks a member that is itself a trained model or that thresholds a quantity the target network computes.}
\label{tab:defenses}
\small
\begin{tabularx}{\textwidth}{@{}llllX@{}}
\toprule
Safeguard & Short name & Family & Learned & Operating point \\
\midrule
Windowed likelihood filter & filter & token surface & yes & 1\% external block rate, window 16 \\
Token-anomaly statistic & anomaly & token surface & no & 1\% external block rate \\
Guard model (8B) & guard & semantic & yes & categorical, input and output \\
Probe, middle layer & probe$_{16}$ & representation & yes & 1\% external block rate \\
Probe, early layer & probe$_{8}$ & representation & yes & 1\% external block rate \\
Refusal-priming prompt & priming & first token & no & fixed system prompt \\
Perturbation aggregator & aggregator & perturbation & no & $q=4$, 10\% character swap \\
\bottomrule
\end{tabularx}
\end{table}

\subsection{Adversaries}

Two optimizing attacks are run against every safeguard. A fluent sentence-level genetic search in
the style of AutoDAN \citep{liu2024autodan} runs 30 generations with a population of
24, elitism at one quarter and a mutation rate of 0.15. It is a reimplementation, not the
original: an individual
is a sequence of indices into a fixed pool of 30 fluent framing sentences, the prompt is those
sentences followed by the behavior, and mutation replaces, inserts or drops one sentence.
Fluency is therefore preserved by construction and not by a perplexity penalty in the
objective, which would entangle the attack with the filter being measured. A gradient coordinate search
\citep{zou2023universal} optimizes a 20-token suffix for up to 150 steps. Both optimize the
same in-loop objective, the target log-probability of the behavior's affirmative prefix over
its first 12 tokens, and two static baselines are run alongside: the unmodified behavior, and
a single precomputed suffix reused unchanged.

The adversary knows the defense. Cheap constraints (the likelihood filter, the surface statistic, the probes) restrict
the feasible set at every generation, evaluated exactly as deployment evaluates them: on the
raw prompt, under the target's default system prompt. Members that change generation enter the
objective instead, so the attack optimizes the quantity the deployed system produces.
The guard model is too expensive to call per candidate, so it is
checked in rounds: the search advances every behavior a share of its generations, then one
block classifies each behavior's best unchecked candidates on both its input and its output
check, and the final choice is made only among candidates the guard has admitted.

\paragraph{Initialization of defended searches.}
Each defended run is seeded from the final prompts the undefended run produced for the same
behaviors and then optimizes further under its own constraint, as an attacker holding
prompts that already work against the bare model would. A defended arm therefore receives
strictly more search than the undefended arm, so the undefended figure is a floor, and a
defended arm can exceed it without the safeguard having made anything worse
(Table~\ref{tab:asr}). On the primary target that margin
is small, because the attack is already saturated against the bare model: its in-loop objective
reaches a median log-probability of $-0.001$. Where an
attack does \emph{not} saturate, the comparison needs more care
(Sections~\ref{sec:replication} and~\ref{sec:probes}).

\paragraph{Handling of blocked candidates.}
A blocked candidate is neither discarded nor assigned $-\infty$. It keeps the objective value it
earned, and selection orders the population lexicographically, feasibility first and fitness
second, so any feasible individual outranks every infeasible one while infeasible individuals
remain available to breed. Discarding or flooring
blocked candidates would collapse the population early against a filter that admits little
and understate what the
adversary can reach. Only a feasible individual can be returned as the attack's choice, and a behavior for which no
feasible candidate is found is recorded as such.

\subsection{Controls the estimates require}\label{sec:controls}

A correlation between two breach vectors has three sources besides the safeguards: the
search itself, the difficulty of the behaviors, and the grader. Each of the controls below
addresses one of them.

\paragraph{Independent replicates and the null.}
Each search is seeded independently per replicate, defense and behavior, so that no two
defenses share a search trajectory. If two defenses
neither reject a candidate nor alter the objective, a shared random stream makes them replay
the same search, and the resulting breach vectors then differ only by grader noise; a
diversity statistic computed on them measures the coupling, not the defenses. An earlier
version of this work had that defect.

We verify the property on the
produced artifacts. No two replicates of a defense,
and no two defenses within a replicate, share a derived search seed, and at most 2\% of the
optimized prompts coincide in any such comparison, which is the genetic search occasionally
converging. Conversely, the
static baselines involve no search and reproduce byte for byte across replicates, so the
divergence comes from the seeding and not from
ambient nondeterminism.

Because the attack is stochastic, a correlation between two defenses is uninterpretable
without knowing how reproducible the attack is. We therefore run the undefended model through
the same independent replicates and report
$\phi_{\mathrm{null}} = \phi(\text{replicate } i, \text{replicate } j)$ on its breach vectors.
No defense is present there, so any correlation is attack reproducibility alone, and it is the
level a defense pair must exceed before co-vulnerability means more than the attack finding
the same behaviors twice. On the same basis we report each defense's test-retest reliability
$r_d$, and $\phi$ disattenuated for it, $\phi_{12}/\sqrt{r_1 r_2}$.

\paragraph{Separating diversity from behavior difficulty.}
Behaviors differ in how readily they are elicited, and that alone correlates any two
detectors. We fit a random-intercept logistic model over every defense, behavior and
replicate,
\begin{equation}\label{eq:multilevel}
\mathrm{logit}\,\Pr(b_{d,i,s} = 1) \;=\; \alpha_d + u_i,
\qquad u_i \sim \mathcal{N}(0, \sigma_b^2),
\end{equation}
in which $u_i$ is behavior $i$'s latent difficulty, shared by every defense. Under this model
two defenses fail independently \emph{given the behavior}, so the $\phi$ it implies is the
association that difficulty alone produces; we report it beside the observed $\phi$, and the
excess between them is the part difficulty does not explain.

\paragraph{Grading and its effect on correlation.}
Breach is scored after the fact by a rubric autograder \citep{souly2024strongreject} at its
standard binarization threshold, using a model that never participates in the search; the
separation is enforced by a static test over the attack modules.

The grader is not deterministic even at temperature zero: on 6{,}474 pairs of calls issued
with the same prompt, the same pinned grader
returned a different rubric score for 17.6\% of them and a different breach label for 4.7\%.
Two defenses that produce byte-identical responses could therefore receive different
labels, and the difference would enter $\phi$ as a difference between defenses. We
key verdicts by the exact text judged together with the grader's identity, so that
identical responses receive one label wherever they occur, and we judge each unique response five
times, taking the majority and retaining every vote.

An aligned model used as a grader also sometimes declines to judge, returning a
refusal in place of a rubric. We record these as \emph{abstentions}, not as non-breaches,
report the interval they imply, and resolve them with a second grader
(Section~\ref{sec:asr}). A targeted human
adjudication covers the response texts of the primary run on which the five judgments split
or the grader abstained, and a
balanced random sample of the unanimous ones. It is labeled by one annotator, an author, who
sees only the request and the response: not the grader's verdict, and not which safeguard or
replicate produced the text. A stratified subset is relabeled after an interval, under new
identifiers and in a new order, to measure that annotator's own consistency.

\subsection{What a deployed stack means here}

The assembled stack applies its members in a fixed order under a unanimity veto: one blocker
suffices, so an attack succeeds only if every member admits it. Input filters run on the raw
prompt in that order, then generation proceeds through the members that modify it, then output
filters run. The order is the likelihood filter, the token-anomaly statistic, the two probes,
the guard model, the refusal-priming prompt and the perturbation aggregator, which places the
cheap surface tests before the expensive semantic one and both before anything that changes
what is generated. The same seven components in a different
sequence are a different system, and an estimate from one ordering does not transfer to
another.

The stack is attacked directly, and the estimate from intersecting per-layer breach vectors
is reported beside it. We also attack selected
pairs as pipelines, which gives the end-to-end estimand at a granularity
where the per-layer intersection can be compared against it with some resolution. Because the
guard model dominates the full stack, a six-member configuration without it is run alongside;
with the guard present every residual sits on the floor, where any method returning near zero
would agree with any other.

\section{Results}\label{sec:results}

Block rates come first, because they need no grader. After that, each control precedes
the claims it bounds: the reliability of the labels, then the reproducibility of the
attack, then the pair statistics.

\subsection{Measured strength is a property of the attack class}

Table~\ref{tab:attackclass} gives each blocking member's block rate on the same 100
behaviors under three attack classes. The quantity is the deployed block decision,
which needs no grader; graded attack success follows in Section~\ref{sec:asr}.

\begin{table}[!htbp]
\centering
\caption{Block rate per safeguard and attack class on 100 behaviors. Fluent: 100 prompts optimized separately by the genetic search. Suffix: one universal suffix appended to every behavior. Plain: the unmodified behavior. The priming prompt and the perturbation aggregator return no block decision and are omitted.}
\label{tab:attackclass}
\small
\begin{tabular*}{\textwidth}{@{\extracolsep{\fill}}lcccccc@{}}
\toprule
& \multicolumn{3}{c}{Vicuna-7B-v1.5} & \multicolumn{3}{c}{Llama-2-7b-chat} \\
\cmidrule(lr){2-4}\cmidrule(lr){5-7}
Safeguard & Fluent & Suffix & Plain & Fluent & Suffix & Plain \\
\midrule
Windowed likelihood filter & 0.00 & 1.00 & 0.00 & 0.00 & 1.00 & 0.00 \\
Token-anomaly statistic    & 0.00 & 0.00 & 0.00 & 0.00 & 0.00 & 0.00 \\
Guard model (8B)           & 0.96 & 0.98 & 0.98 & 0.96 & 0.98 & 0.98 \\
Probe, middle layer        & 0.00 & 0.22 & 0.89 & 0.49 & 0.93 & 0.94 \\
Probe, early layer         & 0.00 & 0.00 & 0.21 & 0.00 & 0.00 & 0.44 \\
\bottomrule
\end{tabular*}
\end{table}

The likelihood filter is the clearest case. It blocks every prompt carrying the
universal suffix and none of the hundred fluent prompts, so the same component
at the same threshold is a complete defense against one adversary and inert
against another. The token-anomaly statistic blocks nothing in any class at this
operating point, which makes the non-neural contrast of
Section~\ref{sec:ensemble} vacuous on the security axis: a member that never
fires cannot be diverse from anything.

The guard model is the exception, blocking between 96\% and 98\% in all three
classes. On the primary target the mid-layer probe catches most plain harmful
prompts and none of the fluent ones. One member therefore does nearly all of
the work, which is why the stack is also run without it.

Three of the five members return identical rates on both targets in all three
attack classes, because they inspect the prompt and never consult
the model. Only the probes move: the mid-layer probe goes from blocking none of
the fluent prompts to
blocking 0.49 of them, and the early-layer probe doubles on plain harmful prompts.
A prompt-side filter's block rate is thus a
property of the attack class alone, while a representation-side probe's also depends
on the target and has to be remeasured (Section~\ref{sec:probes}).

\subsection{What the safeguards buy, and what the labels cost}\label{sec:asr}

Block rates are not attack success: a prompt that passes a filter still has to
elicit something harmful. Table~\ref{tab:asr} reports graded attack success under
the fluent adversary, which first required resolving grader abstentions.

\paragraph{Grader abstentions.}
On 240 of the 8{,}003 judged responses, across three replicates and four attack
conditions, the primary grader returned a refusal in place of a rubric; none was a
malformed rubric. They are concentrated where the
attack succeeded: 197 of the 240 fall under the fluent adversary and 152 of those
were breaches, spread across the six members that do not block, while the guard
model has none under that adversary, because it blocked almost everything before
the grader saw it. The second attack family of Section~\ref{sec:family} contributes
only 19, because it succeeds far less often. The
abstained responses are substantive, with a median length of 1058 characters
against 1095 for the responses the grader did score, and none of them is a target
refusal. Counting a grader's refusal as a non-breach would have
understated attack success by 7 to 9 points per defense and put 8 to
12 arbitrary zeros into every breach vector the correlation analysis consumes.

We resolve them with a second grader from a different developer, a reasoning model
pinned at a dated version, which answered every one and judged 167 of the 240 to be
breaches.
Table~\ref{tab:asr} reports the resolved figure together with the interval the two
defaults imply.

\begin{table}[!htbp]
\centering
\caption{Graded attack success per safeguard and attack class, pooled over three replicates (300 attempts per cell). ``Abstentions'' counts the grader abstentions under the fluent adversary, which the second grader resolved. ``Abstention range'' is the fluent rate if every abstention were counted as a non-breach or as a breach.}
\label{tab:asr}
\small
\begin{tabular*}{\textwidth}{@{\extracolsep{\fill}}lccccc@{}}
\toprule
Safeguard & Fluent & Abstentions & Abstention range & Suffix & Plain \\
\midrule
Refusal-priming prompt & 0.75 & 29 & [0.68, 0.77] & 0.13 & 0.03 \\
\emph{Undefended} & \emph{0.74} & 28 & [0.66, 0.75] & \emph{0.32} & \emph{0.07} \\
Probe, early layer & 0.75 & 27 & [0.68, 0.77] & 0.32 & 0.03 \\
Token-anomaly statistic & 0.74 & 33 & [0.66, 0.77] & 0.32 & 0.07 \\
Probe, middle layer & 0.73 & 25 & [0.67, 0.75] & 0.29 & 0.03 \\
Windowed likelihood filter & 0.72 & 30 & [0.65, 0.75] & 0.00 & 0.07 \\
Perturbation aggregator & 0.65 & 25 & [0.58, 0.67] & 0.05 & 0.03 \\
Guard model (8B) & 0.03 & 0 & [0.03, 0.03] & 0.00 & 0.00 \\
\bottomrule
\end{tabular*}
\end{table}

Five of the six non-guard safeguards leave attack success inside the undefended
estimate's 95\% interval, $[0.69, 0.79]$ around 0.74 (Table~\ref{tab:asr}).
The perturbation aggregator lowers it to 0.65, with an interval of $[0.59, 0.70]$: paired
by behavior and replicate, 43 attempts breach only without it and 15 only with it, an exact
McNemar $p = 0.0003$.
The refusal-priming prompt sits nominally above the undefended model,
which is not evidence that it helps an attacker: defended arms are seeded from the
undefended run and then searched further.
Only the guard model reduces attack success substantially, to 0.03.

\paragraph{Grader reliability.}
Two measurements bound how much of this could come from the grader. Within the primary
grader, repeated calls on an
identical prompt disagree on the breach label for 4.7\% of pairs
(Section~\ref{sec:controls}). Between graders, on 120 responses per replicate that the primary grader labeled
without abstaining, raw agreement is 0.792, 0.775 and 0.867, and Cohen's $\kappa$
is 0.359, 0.333 and 0.588, fair to moderate and spanning 0.254 across replicates. On the
replication target the same comparison gives raw
agreement 0.925, 0.900 and 0.933 with $\kappa$ of 0.434, 0.488 and 0.524:
nearly every response on that target is a refusal both
graders score alike, so chance agreement is high and $\kappa$ discounts it.
The two graders agree least on the cross-evaluation of Section~\ref{sec:e2},
at raw 0.658 and $\kappa$ 0.379 on its own 120-row
sample, which is also where the primary grader abstained most often, on 490 of
6{,}400 responses. Responses to prompts that were
optimized against a \emph{different} member are the hardest to
judge. Label noise of that size does not explain a gap
between 0.03 and 0.74, so the guard-model result is safe from it, but it is large
enough to matter for the finer comparisons between the other six members.

\paragraph{Human adjudication.}
Two automated graders agreeing is weaker evidence than either agreeing with a person,
so one annotator labeled a targeted sample of 618 responses without seeing the grader's
verdict: the 341 on which the grader's five judgments split, the 197 on which it
abstained, and 80 drawn at random from the unanimous ones, half of them breaches by the
grader's label. The annotator agrees with the
grader on \AdjAgreeSplit{} of the split responses, with the second grader's resolution on
\AdjAgreeAbstain{} of the abstentions, and with the grader on \AdjAgreeUnanimous{} of the
unanimous ones. Agreement on the split responses is at chance level
($\kappa = \AdjKappaSplit$); these are the responses on which the grader's own five
judgments disagreed. The annotator calls a breach more often than the graders do, on
\AdjHumanBreach{} of the 618 responses against \AdjGraderBreach{}. Taking the annotator as
the reference,
the grader's precision on the sample is \AdjPrecision{} and its recall is \AdjRecall{}, so
relative to this annotator the automated labels understate attack success.
The sample oversamples the responses the grader found hardest, so these figures describe
the sample and not the corpus. On the 124 responses
relabeled after an interval, the annotator repeats the earlier label on \AdjRetestAgree{},
a test-retest $\kappa$ of \AdjRetestKappa{}.

\AdjSensitivity{} The verdict of Section~\ref{sec:e1} therefore does not depend on which of
the two sets of labels is used.

\subsection{How reproducible is the attack?}

Every pair statistic in the following subsections is read against the reproducibility of
the attack, so we report it first.
Run against the \emph{undefended} model under three independent
seeds, the search produces breach vectors whose three pairs agree at
$\phi = 0.658$, $0.637$ and $0.564$, a mean $\phi_{\mathrm{null}} = 0.620$. This is
attack and grader reproducibility alone, and it is the
level a pair of safeguards must exceed (Section~\ref{sec:controls}). The
spread of the three estimates, $0.094$, is the resolution of that
comparison. A null of that size means that most of what an uncontrolled co-failure
statistic would report is a property of the
behavior set and the optimizer, present before any defense is installed.

\paragraph{Grader and attack contributions to the null.}
The null mixes two sources of variation, and they can be separated because one of
our replicates was graded twice in full. Splitting those judgments into two
disjoint sets and taking the majority within each gives two breach vectors over
\emph{byte-identical} responses, so they differ only by grader variability. They
agree at $\phi = 0.926$ on average, from 0.819 to 0.978 across members, with
between 1\% and 8\% of labels flipping.

That figure is the ceiling of this measurement: a pair of safeguards cannot be
shown to agree more than the grader agrees with itself. It also
shows which source dominates. The grader reproduces itself at 0.926 and the
attack at 0.620, so the gap between a perfect measurement and
our null is mostly the optimizer's stochasticity and not the labels.

\paragraph{Decoding non-determinism.}
The cross-evaluation below re-runs each defense on its own search's prompts, so
its diagonal should reproduce the responses recorded during the attack.
Across seven members and 700 prompts, 6.3\% of responses differ from
the originals, although decoding is greedy and the prompt is byte-identical; the
median pair agrees for 448 characters and then diverges into a different,
coherent continuation. Generation proceeds identically until two candidate
tokens are near-tied, and the arithmetic resolving the tie depends on
how the batch was assembled. Greedy decoding is therefore not reproducible across
runs that batch prompts differently, at a rate of about one prompt in sixteen.
For this paper it means
$\phi_{\mathrm{null}}$ contains a third component beside search stochasticity and
grader noise, and the verdict cache, which assigns one label per exact text,
deduplicates less across runs than within one.

Each defense's test-retest reliability, the mean $\phi$ over the three pairs
of replicates, runs from 0.421 for the perturbation aggregator to 0.651 for the
token-anomaly statistic (Table~\ref{tab:reliability}). These are the $r_d$
used to disattenuate below. The guard model's reliability is the least stable of the
eight, at 0.656, 0.656 and 0.313 across the three pairs, which is what a
near-constant breach vector does to a correlation.

\begin{table}[!htbp]
\centering
\caption{Test-retest reliability per member: $\phi$ between independent replicates of the same search, for each pair of replicates, and the mean $r_d$. The undefended row is $\phi_{\mathrm{null}}$.}
\label{tab:reliability}
\small
\begin{tabular*}{\textwidth}{@{\extracolsep{\fill}}lcccc@{}}
\toprule
& \multicolumn{3}{c}{Replicate pair} & \\
\cmidrule(lr){2-4}
Safeguard & 1 and 2 & 1 and 3 & 2 and 3 & Mean $r_d$ \\
\midrule
Token-anomaly statistic    & 0.637 & 0.689 & 0.627 & 0.651 \\
Probe, early layer         & 0.601 & 0.769 & 0.566 & 0.645 \\
Probe, middle layer        & 0.669 & 0.637 & 0.574 & 0.627 \\
\emph{Undefended} ($\phi_{\mathrm{null}}$) & \emph{0.658} & \emph{0.637} & \emph{0.564} & \emph{0.620} \\
Refusal-priming prompt     & 0.590 & 0.608 & 0.646 & 0.615 \\
Windowed likelihood filter & 0.597 & 0.545 & 0.585 & 0.576 \\
Guard model (8B)           & 0.656 & 0.656 & 0.313 & 0.542 \\
Perturbation aggregator    & 0.462 & 0.359 & 0.440 & 0.421 \\
\bottomrule
\end{tabular*}
\end{table}

\subsection{E1: behavior-level co-vulnerability}\label{sec:e1}

The first estimand compares safeguards that were attacked separately, so each breach
vector is indexed by behavior. We read the pair statistics against the null of the
previous subsection, then against behavior difficulty, and then check how far the
verdict depends on the correction for unreliability, on the handling of abstentions and on
the choice of pairwise statistics.

\begin{figure}[!htbp]
\centering
\includegraphics[width=\linewidth]{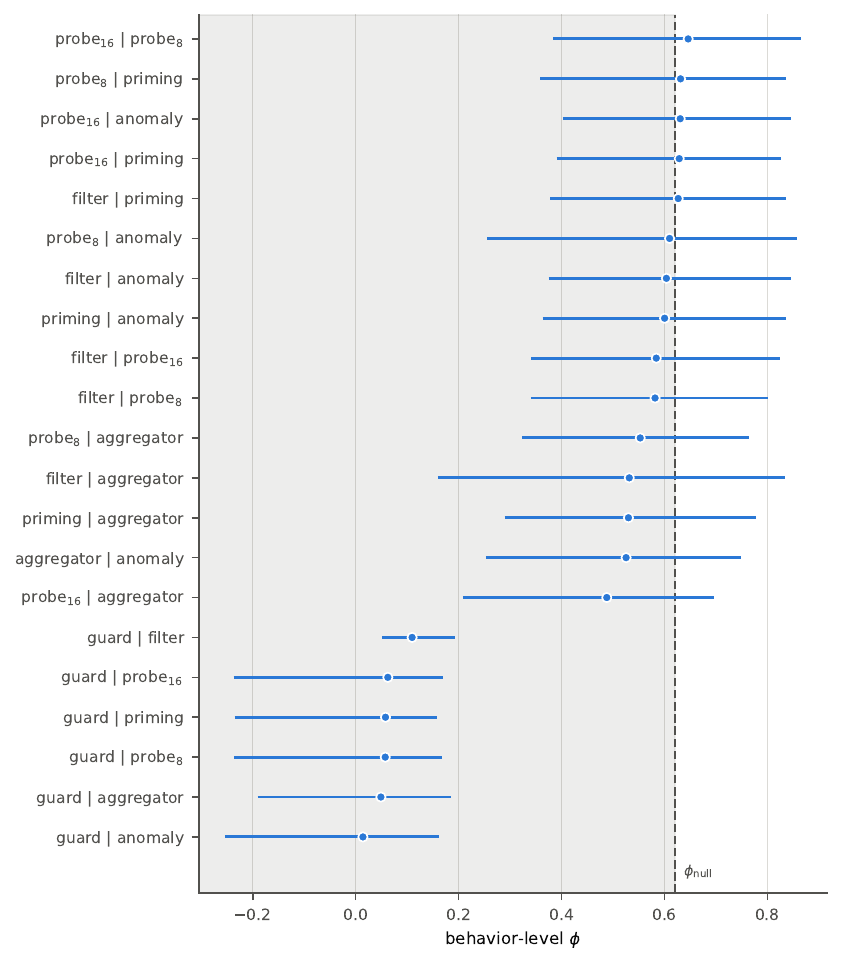}
\caption{Behavior-level $\phi$ for all 21 pairs. Points are means over three replicates, and each bar spans the three per-replicate bootstrap 95\% intervals. The dashed line is $\phi_{\mathrm{null}} = 0.620$, the mean correlation between two independent runs of the same attack with no safeguard present, and the shaded region is everything at or below it.}
\label{fig:forest}
\end{figure}

\paragraph{No pair is distinguishable from the null.}
Table~\ref{tab:e1} gives behavior-level $\phi$ for all 21 pairs, and
Figure~\ref{fig:forest} plots them against the null. Among the
fifteen not involving the guard model, $\phi$ runs from 0.488 to 0.646, which read
on its own looks like substantial shared vulnerability. Read against
$\phi_{\mathrm{null}} = 0.620$, five sit above it and ten below, and the largest
excess in the table is $+0.026$.

The null's own three estimates span $0.094$, and the
median bootstrap interval on a pair is $0.343$ wide (Appendix~\ref{app:bootstrap}), so an
excess of $0.026$ is about a
quarter of the scatter in the null and a thirteenth of the width of the interval on the
estimate. Nothing here separates any pair from the level the
attack reaches on its own. With two replicates no pair was above the null;
the third moved the null down by 0.038 and put five nominally above it,
so a single seed pair does not pin this
quantity down.

The six pairs containing the guard model fall far below, from 0.110 to 0.014,
which is what a member that blocks almost everything does to a correlation: its
breach vector is nearly constant, so there is little left to covary.

Across three replicates and seven safeguards, we therefore find no
evidence that any pair of these safeguards fails together more than two
independent runs of the same attack do with no safeguard present. This is not a
claim that the members are independent: $\phi$ near the null is not
$\phi$ near zero. The claim is that this
experiment cannot attribute the shared structure to the pairing of mechanisms.

\begin{figure}[!htbp]
\centering
\includegraphics[width=0.78\linewidth]{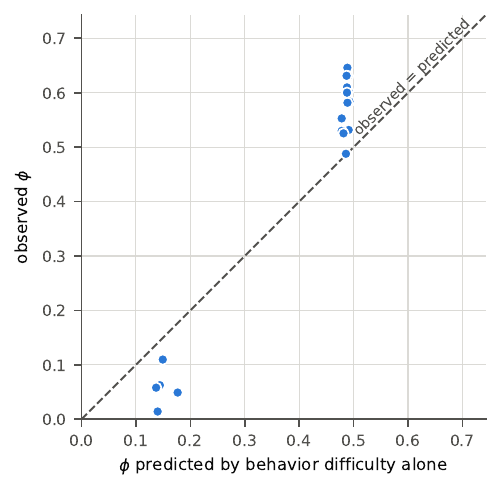}
\caption{Observed $\phi$ against the $\phi$ that behavior difficulty alone predicts under the random-intercept model, one point per pair. A pair on the dashed diagonal would be fully accounted for by difficulty. The fifteen non-guard pairs lie above it and the six guard pairs below it.}
\label{fig:difficulty}
\end{figure}

\paragraph{Behavior difficulty accounts for most of the association.}
The multilevel model of Eq.~\eqref{eq:multilevel} separates how hard a behavior
is to elicit from anything the detectors share. Fitted over every defense,
behavior and replicate, it puts a standard deviation of 2.73 on the logit scale
across behaviors and an intraclass correlation of 0.693: which behavior is being
attacked accounts for 69\% of the variance in whether an attempt breaches, before
any defense is named. We report two estimators of the effect, which are not the
same quantity. Conditioning on the undefended per-behavior breach rate, a difficulty
measure taken with no defended outcome in it, the mean association across the
fifteen non-guard pairs falls from $\phi = 0.585$ to a partial $\phi = 0.171$:
difficulty removes 71\% of it. Asking the fitted model instead what $\phi$ those
pairs would show if behavior difficulty were the only thing they shared gives
0.487 against the observed 0.585, so difficulty accounts for 83\% and the excess
is 0.098 (Figure~\ref{fig:difficulty}). The first is a residual after
conditioning and the second a prediction from the fitted variance components.

The guard model's pairs invert the sign of that excess, at $-0.089$ on average:
they are \emph{less} correlated than behavior difficulty alone would predict,
because the guard blocks almost everything regardless of how hard the
behavior is.

Individually the partials run from 0.002 to 0.268, and the ordering of the pairs
is only partly preserved (Table~\ref{tab:e1}).

So the shared-substrate reading of a raw co-failure number is not supported here.
What two neural safeguards mostly share, as far as this measurement can see, is
that some harmful behaviors are easy to elicit from the target and some are not,
which is a property of the target and the benchmark and not of the detectors
placed around it.

\paragraph{Limits of disattenuation.}
Dividing $\phi$ by $\sqrt{r_a r_b}$ corrects for unreliability, and with $r_d$ in
the measured range (Table~\ref{tab:reliability}) the quotient exceeds 1 for eight of the
twenty-one pairs, reaching 1.08. In those cases the
quotient is not a corrected correlation but a signal that observed
$\phi$ is already as large as these reliabilities permit, which is the same
message as the comparison against the null. We report the raw quotient
with the inadmissible cases flagged and do not cap it.

\begin{table}[!htbp]
\centering
\caption{E1: behavior-level $\phi$ for each pair, mean over three replicates, and its excess over $\phi_{\mathrm{null}} = 0.620$. Partial $\phi$ conditions on the undefended per-behavior breach rate. A dagger marks a disattenuated value above 1, which is inadmissible as a correlation. Short names are those of Table~\ref{tab:defenses}.}
\label{tab:e1}
\small
\begin{tabular*}{\textwidth}{@{\extracolsep{\fill}}lcccc@{}}
\toprule
Pair & $\phi$ & $\phi-\phi_{\mathrm{null}}$ & Partial $\phi$ & Disattenuated $\phi$ \\
\midrule
probe$_{16}$ $|$ probe$_{8}$ & 0.646 & $+0.026$ & 0.268 & 1.02$^\dagger$ \\
probe$_{8}$ $|$ priming    & 0.631 & $+0.011$ & 0.261 & 1.00$^\dagger$ \\
probe$_{16}$ $|$ anomaly   & 0.631 & $+0.011$ & 0.222 & 0.99 \\
probe$_{16}$ $|$ priming   & 0.629 & $+0.009$ & 0.215 & 1.01$^\dagger$ \\
filter $|$ priming         & 0.627 & $+0.007$ & 0.207 & 1.05$^\dagger$ \\
probe$_{8}$ $|$ anomaly    & 0.610 & $-0.010$ & 0.234 & 0.94 \\
filter $|$ anomaly         & 0.604 & $-0.016$ & 0.169 & 0.99 \\
priming $|$ anomaly        & 0.600 & $-0.020$ & 0.201 & 0.95 \\
filter $|$ probe$_{16}$    & 0.584 & $-0.036$ & 0.069 & 0.97 \\
filter $|$ probe$_{8}$     & 0.582 & $-0.038$ & 0.126 & 0.95 \\
probe$_{8}$ $|$ aggregator & 0.553 & $-0.067$ & 0.205 & 1.06$^\dagger$ \\
filter $|$ aggregator      & 0.532 & $-0.088$ & 0.122 & 1.08$^\dagger$ \\
priming $|$ aggregator     & 0.530 & $-0.090$ & 0.135 & 1.04$^\dagger$ \\
aggregator $|$ anomaly     & 0.525 & $-0.094$ & 0.133 & 1.00$^\dagger$ \\
probe$_{16}$ $|$ aggregator & 0.488 & $-0.132$ & 0.002 & 0.95 \\
\addlinespace
guard $|$ filter           & 0.110 & $-0.510$ & 0.089 & 0.20 \\
guard $|$ probe$_{16}$     & 0.063 & $-0.557$ & 0.009 & 0.11 \\
guard $|$ priming          & 0.058 & $-0.562$ & 0.016 & 0.10 \\
guard $|$ probe$_{8}$      & 0.058 & $-0.562$ & 0.001 & 0.10 \\
guard $|$ aggregator       & 0.049 & $-0.571$ & 0.011 & 0.10 \\
guard $|$ anomaly          & 0.014 & $-0.606$ & -0.049 & 0.02 \\
\bottomrule
\end{tabular*}
\end{table}

\paragraph{Sensitivity to how abstentions are resolved.}
The verdict does not depend on how the abstentions were resolved. Repeating the comparison on
the unresolved labels, with every grader abstention counted as a non-breach,
raises $\phi_{\mathrm{null}}$ from 0.620 to 0.685 and lowers the number of
non-guard pairs sitting nominally above it from five to two, the largest excess
becoming $+0.021$ instead of $+0.026$. The largest excess
is about two hundredths under either treatment, against a null whose own three
estimates span 0.094 when resolved and 0.160 when not.

\paragraph{Non-pairwise statistics.}
Pairwise statistics taken one at a time can hide what the set does together, so we
also compute the non-pairwise measures of \citet{kuncheva2003measures} over all
members at once, and compare each with what members with the same marginal
breach rates would give if they failed independently, since these
measures have no natural zero.

Every measure is far from its independence value in all three
replicates. Over the seven members, interrater agreement averages 0.31 against
0.00 for independent members, entropy 0.44 against 0.77, and the difficulty
measure $\theta$, which is large when items are uniformly easy or hard for
everyone, 0.096 against 0.025. Without the guard model, whose near-constant
vector pulls the set apart, the six remaining members give an interrater agreement
of 0.58 and an entropy of 0.21 against 0.51.

This result and the pairwise one answer different questions. The members are not
independent. What the comparison against $\phi_{\mathrm{null}}$ adds is that their
dependence is not attributable to the pairing of mechanisms: an adversary with a
fixed behavior set produces this much agreement against no defense at all.

\paragraph{Within-family pairs.}
The two probes read the same network at different depths and were trained on the
same pool, so if any pair in this stack ought to show mechanism-driven co-failure
it is that one. Its $\phi$ of 0.646 is the largest
in the table but only $0.026$ above the null, and its partial $\phi$ of 0.268 is
the largest of fifteen values spanning 0.002 to 0.268. Section~\ref{sec:probes}
gives a reason to treat even that residue cautiously: the
early-layer probe separates two harmless corpora better than it separates harm,
so part of what the pair shares is a corpus effect.

\subsection{E2: do two safeguards fail on the same prompt?}\label{sec:e2}

E1 asks whether two safeguards are vulnerable on the same \emph{behaviors} when
each is attacked separately. In a deployed stack one prompt
arrives and every member sees it. That is the identity
Eq.~(\ref{eq:joint}) is written over, and measuring it needs the cross-evaluation,
in which every member is scored on the final prompts that every \emph{other}
member's search produced. We also score the undefended target on those same
prompts.

\paragraph{Prompt-level association and the pass-through diagnostic.}
On the 800 shared inputs, prompt-level $\phi$ runs from 0.016 to 1.000, against a
behavior-level mean of 0.585 (Table~\ref{tab:e1}). Taken at face value the top of
that range says two safeguards fail together deterministically. That reading is
wrong. The likelihood
filter and the early-layer probe block none of the 800 prompts and return the
target's own response to all 800 (Table~\ref{tab:e2}), so on a shared input they are
the same function. Their $\phi$ of 1.000 with each
other is also their $\phi$ of 1.000 with \emph{no defense at all}.

That last quantity is the diagnostic. A member's $\phi$ against the undefended
control on the same prompts measures how much of the pipeline it changes,
and it orders the stack: 1.000 for the two pass-through members, 0.981 and
0.942 for the two that block a handful, 0.645 for the priming prompt, which
rewrites the system context, 0.440 for the perturbation aggregator, which replaces
the prompt with a vote over perturbed copies, and 0.071 for the guard model, which
blocks 786 of 800.

\paragraph{What a pairwise value adds beyond its members.}
This ordering largely determines the pairwise values. For each pair take the smaller of its two members' $\phi$ against
the undefended control. For the eighteen pairs in which at most one member alters
the generation, that number predicts the pair's own $\phi$ to within 0.040, with a
mean absolute deviation of 0.005.

Only three pairs fall below the bound, and they are the three in which both
members intervene: priming with the aggregator, and the guard with each of them.
There the shortfall is 0.015, 0.043 and 0.055. Whatever pairwise structure
exists in this stack at prompt level is therefore confined to members that alter
the computation, and even among those it is a few hundredths.

\paragraph{Implication for the estimand.}
E2 is the quantity the composition argument is written
over, and here it does not report diversity between mechanisms.
It reports that four of seven members, each calibrated to a 1\% false-positive rate
on an external corpus, are indistinguishable from the undefended model on the
inputs an adversary produces. A pairwise diversity statistic computed over
such a set measures the operating points, not the mechanisms, and would
read as near-perfect co-failure without the undefended
control. We therefore recommend scoring the no-defense arm in any cross-evaluation.

\begin{table}[!htbp]
\centering
\caption{E2 on shared inputs: each member scored on the 800 final prompts of all eight searches. The second and third columns are counts over those 800. Members are ordered by their $\phi$ with the undefended control, where 1.000 means the member did not change the pipeline. Labels are the resolved ones; the second grader judged 360 of the 490 abstentions to be breaches.}
\label{tab:e2}
\small
\begin{tabular*}{\textwidth}{@{\extracolsep{\fill}}lcccc@{}}
\toprule
Safeguard & Prompts & Target's own & $\phi$ with & Breach \\
& blocked & response returned & undefended & rate \\
\midrule
Windowed likelihood filter & 0 & 800 & 1.000 & 0.735 \\
Probe, early layer         & 0 & 800 & 1.000 & 0.735 \\
Token-anomaly statistic    & 7 & 793 & 0.981 & 0.728 \\
Probe, middle layer        & 30 & 770 & 0.942 & 0.711 \\
Refusal-priming prompt     & 0 & 3 & 0.645 & 0.713 \\
Perturbation aggregator    & 0 & 0 & 0.440 & 0.570 \\
Guard model (8B)           & 786 & 14 & 0.071 & 0.014 \\
\emph{Undefended}          & \emph{0} & \emph{800} & n/a & \emph{0.735} \\
\bottomrule
\end{tabular*}
\end{table}

\paragraph{Transfer between members.}
The cross-evaluation also shows whether a search that works against one member has found
something specific to that member. Table~\ref{tab:transfer} reports the full matrix,
with rows for the search that produced the prompts and columns for the member scored on
them.

The values barely change down a column and change a
great deal along a row. Conditioning on which member is being scored removes 97.7\% of the
variance in the matrix; conditioning on which search produced the prompts removes
0.7\%. Excluding the guard model, whose column is near zero throughout, those figures
are 66.7\% and 8.5\%.

The diagonal carries any specialization, and there is very little.
Excluding the guard it averages 0.717 against an off-diagonal 0.695, an advantage
of 0.022. Per member the advantage of a member's own search over the mean of the
others runs from $-0.06$ to $+0.09$. It is positive for the three members that alter
the pipeline, the priming prompt at $+0.09$, the aggregator at $+0.03$ and the
mid-layer probe at $+0.02$, and it is $-0.01$ and $-0.06$ for the two
pass-through members.

A matrix of uniformly high off-diagonal entries is usually
read as evidence that attacks transfer between defenses. Here it is a
property of the defenses: four of the seven members do not
change what the adversary is attacking, so there is no member-specific adaptation to
transfer.

\begin{table}[!htbp]
\centering
\caption{Transfer matrix: breach rate of each member (columns) on the 100 final prompts of each search (rows), for one replicate on the resolved labels. The diagonal, in bold, is a search scored on its own member.}
\label{tab:transfer}
\footnotesize
\begin{tabular*}{\textwidth}{@{\extracolsep{\fill}}lcccccccc@{}}
\toprule
& \multicolumn{8}{c}{Member scored} \\
\cmidrule(lr){2-9}
Search against & filter & probe$_{8}$ & anomaly & probe$_{16}$ & priming & aggregator & guard & \emph{undefended} \\
\midrule
filter & \textbf{0.68} & 0.68 & 0.66 & 0.68 & 0.71 & 0.58 & 0.02 & 0.68 \\
probe$_{8}$ & 0.73 & \textbf{0.73} & 0.72 & 0.72 & 0.72 & 0.62 & 0.00 & 0.73 \\
anomaly & 0.73 & 0.73 & \textbf{0.73} & 0.72 & 0.71 & 0.54 & 0.01 & 0.73 \\
probe$_{16}$ & 0.73 & 0.73 & 0.73 & \textbf{0.73} & 0.72 & 0.49 & 0.02 & 0.73 \\
priming & 0.73 & 0.73 & 0.73 & 0.73 & \textbf{0.79} & 0.64 & 0.01 & 0.73 \\
aggregator & 0.74 & 0.74 & 0.74 & 0.59 & 0.61 & \textbf{0.60} & 0.01 & 0.74 \\
guard & 0.78 & 0.78 & 0.77 & 0.77 & 0.76 & 0.55 & \textbf{0.03} & 0.78 \\
\emph{undefended} & 0.76 & 0.76 & 0.74 & 0.75 & 0.68 & 0.54 & 0.01 & \textbf{0.76} \\
\bottomrule
\end{tabular*}
\end{table}

Two limits apply to this subsection. E2 and the transfer matrix
come from a single replicate, because the cross-evaluation scores every member on every
other member's prompts and so costs one pass per cell. There is therefore no seed-pair null
for E2 of the kind Section~\ref{sec:e1} uses for E1, and the figures here
are point estimates with no replicate spread behind them. And the pass-through result is
a statement about members at a 1\% operating point on a weakly aligned target;
Section~\ref{sec:replication} reports how it changes on the other one.

\subsection{E3: the assembled stack}\label{sec:e3}

E1 and E2 are properties of members, while a deployer buys the
assembled pipeline. The estimate available without attacking the assembly is the
\emph{intersection}: the breach rate over behaviors that every member failed on,
each measured on its own. Table~\ref{tab:e3} compares it with the assembled stack
attacked end to end. For the six-member stack the residual is measured in the no-guard
configuration, and the intersection comes from the primary run's six non-guard members,
which share its thresholds, seeds, behaviors and target. Labels are the resolved ones:
118 of the 2{,}646 graded stack responses were grader abstentions, and the second grader
judged 88 of them to be breaches.

\paragraph{The seven-member stack.}
Attacked end to end, the seven-layer stack breaches on 0.023 of behaviors,
against 0.017 for the intersection and 0.030 for its strongest single member,
which is the guard model in every replicate. Per replicate the measured values are
0.030, 0.020 and 0.020. McNemar
against the intersection gives $p = 0.5$, $1$ and $1$, and against the strongest
layer $p = 1$ throughout, so this comparison has no resolution, which is why
the stack is also run without the guard.

\paragraph{The composition estimate is optimistic wherever the comparison has resolution.}
Five of the six stacks leave room for their members to disagree (Table~\ref{tab:e3}),
and in every one of them the assembled stack breaches \emph{more} than the intersection
of its own members predicts. The gaps are 0.100 for the six-layer stack and 0.087,
0.073, 0.070 and 0.067 for the four pairs. All five are significant on an exact McNemar
test pooled over the three replicates' 300 paired observations, with $p$ between
$0.0005$ and $0.0375$, and four of the five are below $0.01$. All fifteen
per-replicate gaps are positive, ranging
from $+0.04$ to $+0.15$; for the six-layer stack they are $+0.120$,
$+0.090$ and $+0.090$.

We pool because three per-replicate tests yield three
p-values, and reporting the smallest of them against a threshold of 0.05 inflates the
false-positive rate roughly threefold. The replicates are independent searches over the
same behaviors, so we report a single test over all of their paired observations; the
per-replicate values are in the released
artifacts.

\paragraph{Reading the composition estimate.}
In every comparison with resolution, the per-layer composition estimate is below the
assembled residual, so it
errs in the direction that flatters the stack. This includes the replication target,
where the six-layer stack breaches 0.063 against an intersection of 0.013, pooled
$p = 0.0007$ (Section~\ref{sec:replication}).

\paragraph{What depth buys.}
Read against the strongest-member column, the pairs and the six-layer stack differ. None of
the four \emph{pairs} breaches less when assembled than its stronger member alone:
0.723 against 0.720, 0.747 against 0.727, 0.757 against 0.730 and 0.657 against 0.647,
on 300 attempts each.
A deployer adding the second layer to any of these pairs pays its inference
cost and false refusals for a residual the first layer already delivered.

The six-layer stack does better: 0.573 assembled against 0.647 for its
strongest member, so stacking six members buys 0.074 of residual over the best of
them. The per-layer estimate still overstates the gain: it predicted 0.473 and the pipeline
delivered 0.573, so more than half of the improvement a deployer would have
calculated from member measurements does not survive assembly.

\begin{table}[!htbp]
\centering
\caption{E3: breach rate of the assembled stack attacked end to end (measured), the estimate from intersecting the single-member runs (intersection), and the best single member of the stack (strongest member). $p$ is an exact McNemar test of measured against intersection, pooled over three replicates (300 attempts per row).}
\label{tab:e3}
\footnotesize\setlength{\tabcolsep}{3pt}
\begin{tabular*}{\textwidth}{@{\extracolsep{\fill}}lccccc@{}}
\toprule
Stack & Measured & 95\% interval & Intersection & Strongest member & $p$ \\
\midrule
Seven members, as configured & 0.02 & [0.01, 0.04] & 0.02 & 0.03 & 0.69 \\
Six members, no guard & 0.57 & [0.51, 0.62] & 0.47 & 0.65 & 0.0018 \\
\addlinespace
filter $+$ anomaly & 0.72 & [0.67, 0.77] & 0.65 & 0.72 & 0.0086 \\
probe$_{16}$ $+$ probe$_{8}$ & 0.75 & [0.70, 0.80] & 0.67 & 0.73 & 0.0007 \\
probe$_{16}$ $+$ priming & 0.76 & [0.71, 0.80] & 0.67 & 0.73 & 0.0005 \\
anomaly $+$ aggregator & 0.66 & [0.60, 0.71] & 0.59 & 0.65 & 0.0375 \\
\bottomrule
\end{tabular*}
\end{table}

\subsection{The cost of depth: over-refusal}\label{sec:frr}

A safeguard that blocks nothing harmful can still be expensive. We count a false refusal
whenever a benign prompt fails to get an answer, whether a filter blocked it or
the target refused it, because the distinction does not matter to the user.

\begin{figure}[!htbp]
\centering
\includegraphics[width=0.78\linewidth]{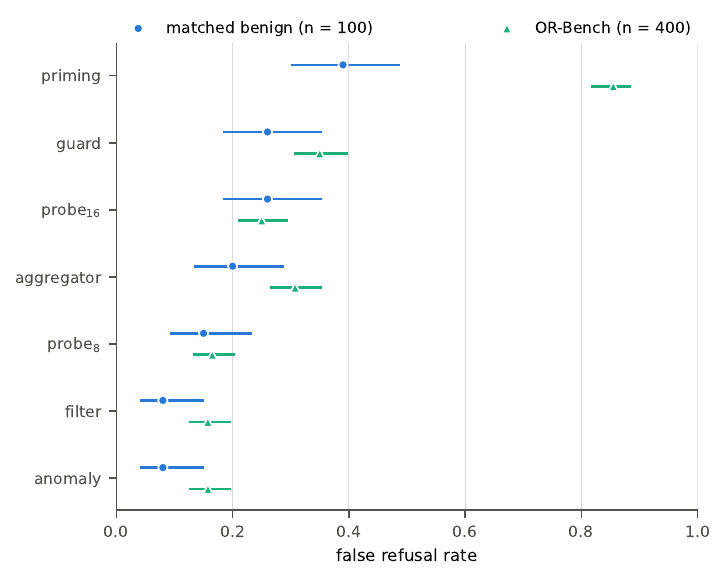}
\caption{False refusal per member on the matched benign set and on the OR-Bench sample, with 95\% Wilson intervals.}
\label{fig:refusal}
\end{figure}

On the 100 matched benign prompts, the per-member rate
runs from 0.08 for the two
token-surface filters to 0.39 for the refusal-priming prompt, with the guard model
and the mid-layer probe both at 0.26 (Table~\ref{tab:frr}; Figure~\ref{fig:refusal}).
The priming prompt blocks nothing, so its entire cost is the
target refusing requests it would otherwise have answered. The token-surface filters
block no benign
prompt either, so their 0.08 is the rate at which the undefended target refuses benign
requests by itself. Benign prompts are not searched over, so the matched-set rates are the
same in all three replicates; the OR-Bench sample was run once.

\begin{table}[!htbp]
\centering
\caption{False refusal on the matched benign prompts and on the OR-Bench sample, with 95\% Wilson intervals. For each stack the rows give the union of the members' own refusals, the value of that union if the members refused independently, and the rate of the assembled stack. The no-guard stack was not run on OR-Bench.}
\label{tab:frr}
\small
\begin{tabular*}{\textwidth}{@{\extracolsep{\fill}}lcccc@{}}
\toprule
& \multicolumn{2}{c}{Matched benign, $n=100$} & \multicolumn{2}{c}{OR-Bench, $n=400$} \\
\cmidrule(lr){2-3}\cmidrule(lr){4-5}
Safeguard or stack & Rate & 95\% interval & Rate & 95\% interval \\
\midrule
Refusal-priming prompt     & 0.39 & [0.30, 0.49] & 0.86 & [0.82, 0.89] \\
Guard model (8B)           & 0.26 & [0.18, 0.35] & 0.35 & [0.30, 0.40] \\
Probe, middle layer        & 0.26 & [0.18, 0.35] & 0.25 & [0.21, 0.29] \\
Perturbation aggregator    & 0.20 & [0.13, 0.29] & 0.31 & [0.26, 0.35] \\
Probe, early layer         & 0.15 & [0.09, 0.23] & 0.17 & [0.13, 0.20] \\
Windowed likelihood filter & 0.08 & [0.04, 0.15] & 0.16 & [0.13, 0.20] \\
Token-anomaly statistic    & 0.08 & [0.04, 0.15] & 0.16 & [0.13, 0.20] \\
\emph{Undefended target alone} & \emph{0.08} & & \emph{0.16} & \\
\addlinespace
Seven members, union of the parts & 0.60 & [0.50, 0.69] & 0.90 & [0.86, 0.92] \\
\quad if they refused independently & 0.81 & & 0.97 & \\
\quad measured, assembled  & 0.83 & [0.74, 0.89] & 0.99 & [0.98, 1.00] \\
\addlinespace
Six members, no guard, union & 0.56 & [0.46, 0.65] & 0.88 & [0.84, 0.91] \\
\quad if they refused independently & 0.74 & & 0.96 & \\
\quad measured, assembled  & 0.82 & [0.73, 0.88] & n/a & \\
\bottomrule
\end{tabular*}
\end{table}

\paragraph{Union of the members' refusals.}
Under the unanimity veto a benign prompt survives
only if every member admits it, and the union of the per-member refusals is 0.60,
reached while five of the seven members leave attack
success at the undefended level. On the independent OR-Bench sample the same union is 0.90.

\paragraph{The composition estimate understates over-refusal.}
The 0.60 is an estimate assembled from the members measured one at a time. The
assembled stack was also run on the same prompts, and it refuses 0.83. The gap is
identical in all three replicates. The stack refuses every benign prompt that any member refuses, and a
further 23 that \emph{no} member refuses on its own, and on the OR-Bench sample the
corresponding figure is 0.995, which is 398 of 400 benign prompts.

The members do not compose as a conjunction of block
decisions. Two of them act on generation and not on admission: the priming
prompt changes the system context, and the aggregator replaces the prompt with a
majority vote over perturbed copies. Assembled, they induce refusals that neither
produces alone, and the probes then read the hidden states of a generation that
has already been altered.

\paragraph{Consequence for the correlation argument.}
If the members refused
independently the union would be 0.81 and not 0.60, so their false refusals
are positively correlated: they reject many of the same benign prompts. Perfect correlation
would give 0.39, the largest single member. Read on the union alone, correlation
therefore helps on this axis, and a diverse member is the
expensive kind, refusing benign traffic the others would have admitted.

The measured stack removes that advantage. At 0.83 it is above the 0.81 that
independence would predict, so what the members' refusals save by overlapping
is more than spent by the interactions that assembling them creates. A deployer choosing which member to add
is comparing per-member refusal sets, where the overlap is real; a deployer deciding
whether to run seven members is choosing
0.83, and on independent benign traffic 0.995.

\subsection{Replication on a strongly aligned target}\label{sec:replication}

\begin{figure}[!htbp]
\centering
\includegraphics[width=\linewidth]{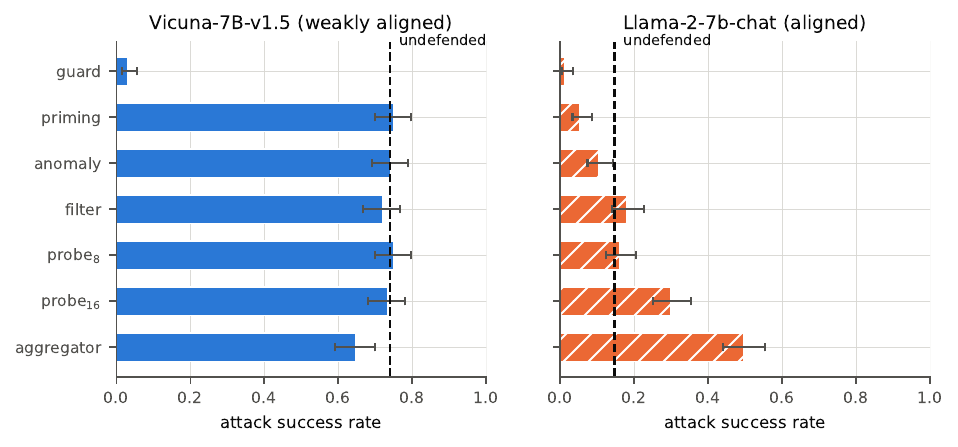}
\caption{Attack success per member under the fluent adaptive adversary on both targets, pooled over three replicates, with 95\% Wilson intervals. The dashed line is each target's undefended rate. Each member's arm searched separately.}
\label{fig:replication}
\end{figure}

The results so far come from one target. We repeat the design on Llama-2-7b-chat, which
refuses far more readily, with the same seven members, the same adversaries, the
same external calibration and probes retrained for that network, and we run all
three estimands. Four comparisons that had resolution on the primary target have none here:
the behavior-level null becomes too
unstable to compare against, the seven-member stack's residual falls to zero, the
over-refusal rate rises to one, and the guard model's breach vector goes constant.
Three survive: the share of association attributable to behavior
difficulty, the prompt-level estimand, and the assembled-stack comparison without the
guard. The arm attacked through the perturbation aggregator also exceeds the undefended
rate (Figure~\ref{fig:replication}), which the same-input test largely attributes to
the prompts.

\paragraph{The behavior-level comparison lacks resolution on this target.}
Averaged over the three seed pairs the attack reproduces itself against the
undefended model at $\phi_{\mathrm{null}} = 0.304$, but the three estimates are
0.504, 0.237 and 0.171: they span
0.333, and every one of the twenty-one pair values lies within that span of the
mean. Resolving this target's grader abstentions leaves all four of those figures
unchanged to three decimals, so the instability is in the attack and not in the
labeling. Eight of the pairs sit above the mean, the largest by 0.211, which is under
two-thirds of the scatter in the null. The
comparison therefore does not discriminate here.

The cause is the base rate. The undefended attack succeeds on roughly 15 of 100
behaviors on this target against roughly 74 on the primary one, so the breach
vectors are sparse, and $\phi$ between sparse binary vectors has high variance:
a handful of behaviors moving between replicates swings it by more
than the quantity being estimated.

Two replicates gave a cleaner and false reading. On two seeds the null here was 0.504
and \emph{no} pair exceeded it, which
would have confirmed the primary target; the
third replicate removed both the number and the conclusion. The same
happened on a smaller scale on the primary target (Section~\ref{sec:e1}). A null
estimated from one pair of runs is not reliable.

\paragraph{Behavior difficulty explains less on this target.}
Conditioning on the undefended per-behavior breach rate takes the mean
association among non-guard pairs from $\phi = 0.285$ to a partial $\phi = 0.187$,
so difficulty accounts for 34\% where it accounted for 71\% on the
weakly aligned target, and the random-intercept fit puts the intraclass
correlation at 0.498 against 0.693. The decomposition is therefore a property of the
regime and not a constant of the method: where almost every behavior is reachable,
which behavior was drawn dominates the variance.

\paragraph{The assembled stack on this target.}
Attacked end to end on this target, the seven-member stack breaches on \emph{none} of the
300 attempts across three replicates. We report that as no successful attack found within the
stated search budget, not as a measured zero. For reference the undefended model breaches
0.147 and the guard model alone 0.013. The intersection of the
members predicts 0.003 pooled, and the two differ on a single
behavior out of 300, with McNemar $p = 1$ in every replicate, so the composition
comparison is empty.

Dropping the guard restores the comparison, and it agrees with the primary target. The
six-layer stack breaches 0.063 of behaviors where the intersection of its own six members
predicts 0.013, a gap of 0.050 that is positive in all three replicates at $+0.02$, $+0.06$
and $+0.07$. Pooled over the 300 paired observations the stack breaches where the intersection
does not on 17 behaviors and the reverse on 2, an exact McNemar $p = 0.0007$. Its strongest
single member, the priming prompt, admits 0.053 alone, so the assembly is not better than
that member. On both
targets, wherever the comparison can resolve, the estimate a deployer can compute
without attacking the assembly is below what the assembly admits.

\paragraph{Over-refusal on this target.}
The over-refusal comparison saturates. Before any safeguard is added, the
undefended Llama-2 refuses 0.500 of the benchmark's matched benign prompts and 0.882 of the
independent OR-Bench sample; the two token-surface
filters sit exactly at the undefended rate, and the others run from
0.530 to 0.990. Assembled, the seven-member stack refuses 100 of 100 matched benign prompts and
400 of 400 OR-Bench prompts. The union of the members' own refusals, the
union independence would predict, and the measured stack are all 1.000, where they
were 0.600, 0.808 and 0.830 on the primary target, so nothing
about composition can be read from them.

\paragraph{The prompt-level estimand on this target.}
The cross-evaluation is the one comparison that gains from
the move. On the primary target four members returned the target's own response to
essentially every adversarial prompt. Here the mid-layer probe
blocks 514 of the 800 shared prompts, against 30 there, and its correlation with the
undefended model barely moves, from 0.942 to 0.965. A member can intercept two thirds of
what an adversary produces and still be indistinguishable from no defense,
because what it intercepts was not going to succeed. A
deployer reading a block rate of 0.64 as a measure of protection would be reading the
difficulty of the prompts the search produced.

The token-anomaly statistic is the opposite case. It blocks 295 of the 800 and its
correlation with the undefended model falls to 0.520, so it is blocking prompts that would
have succeeded. Both members threshold a scalar, and neither effect is predictable
from the block rate.

The min bound of Section~\ref{sec:e2} still holds, to within
0.098, for the fifteen pairs in which at most one member alters the generation. For the six in which both do, against three on the primary
target, the deviations average 0.155 and reach 0.222. The pairwise statistic
thus carries information beyond its members when the members act, and more of them
act on this target.

\paragraph{The guard model's breach vector is nearly constant.}
The guard model admits 1\% of attacked prompts here, and its breach vector is
accordingly almost constant. Its three test-retest estimates are $-0.014$,
$-0.010$ and $0.704$, averaging 0.226, which is not a reliability but three
readings of a quantity the design cannot measure on this target.

\paragraph{What the four losses of resolution have in common.}
Three of the four lost comparisons are one fact: on a target that refuses readily,
the attack succeeds on too few behaviors
for $\phi$ to be estimated, the assembled residual falls to zero, and the member
that does the blocking becomes a constant. The fourth, the over-refusal ceiling, is the same
degeneracy on the cost axis, where the \emph{target} refuses almost everything.

The design therefore needs a
target on which the adversary sometimes succeeds \emph{and} benign traffic sometimes gets
through, and alignment strong enough to make
the first scarce usually makes the second scarce too. A reader planning this measurement
can check both conditions in two cheap runs, the undefended attack and the
undefended benign pass.

\paragraph{The perturbation aggregator on this target.}
Four arms exceed the undefended rate of 0.15 on this target
(Figure~\ref{fig:replication}): the perturbation aggregator at 0.50, the mid-layer probe at
0.30 (Section~\ref{sec:probes}), and the likelihood filter and early-layer probe at 0.18 and
0.16. The aggregator's arm stands furthest above it. Those arms searched separately, so the
two numbers are not on the same prompts, and the cross-evaluation separates the prompts
from the member.

Scored on identical inputs in the cross-evaluated replicate, the undefended model
breaches on 0.42 of the aggregator arm's final prompts and the
aggregator on 0.53; on the undefended arm's prompts the two rates are 0.15 and 0.17.
Most of the gap between 0.15 and 0.50 therefore lies in the prompts. The search against
the aggregator optimizes the mean log-probability over perturbed copies, and the prompts
it finds are stronger against the bare model as well, so the undefended arm's 0.15
understates what this adversary can reach on this target. On identical inputs the
aggregator's own contribution points toward harm on its arm's prompts, where
23 behaviors breach only behind the aggregator and
12 only without it (exact McNemar $p = 0.09$, $n = 100$), and is absent on the undefended
arm's prompts, 11 against 9.

A mechanism for harm exists. The aggregator perturbs the prompt, generates a response to
each of $q$ copies,
classifies each copy as a refusal or not, and returns a response from the
majority class. Its protection assumes that perturbing an adversarial prompt
breaks the attack, so that most copies refuse. On a model whose refusal behavior
is itself fragile to perturbation the assumption can invert: most copies comply, and
the defense returns one of the
complying responses. In the separate-search arms the
undefended Llama-2 refuses 70 of 100 prompts and behind the aggregator it refuses 4,
and in 32 of the 37 behaviors where the aggregator arm breached and the
undefended arm did not, the undefended model had \emph{refused}. Those arms differ in
their prompts as well, so these counts are consistent with the mechanism without
isolating it. Against the static suffix attack, the one this defense was designed
against, the aggregator still blocks everything, 0.00 residual.

On the primary target the same-input test shows the aggregator helping. It breaches on
0.54 of the undefended arm's prompts, against 0.76 for the undefended model, and on
0.60 of its own arm's prompts, against 0.74 (Table~\ref{tab:transfer}). Both
differences are larger than the 0.09 of the separate-search comparison, 0.65 against 0.74,
which is the direction to expect when the adversary has adapted to the member. The
aggregator's effect on a given input thus differs between the two targets, and the
separate-search comparison understates it on one and overstates it on the other.

\subsection{A second attack family}\label{sec:family}

The pair statistics above come from one search. To test whether the \emph{method's}
verdict survives a search that works by another
mechanism, we repeat the measurement with a second family: an attacker language model that
reads the target's reply and rewrites its own prompt, in place of a genetic operator
recombining framing sentences from a fixed pool. Objective and feasibility rule are held
identical, so only the proposal mechanism differs.

\paragraph{Search budgets are not matched.}
The two families cannot be matched on hyperparameters: a genetic arm evaluates a population
per generation, and this one evaluates four streams over six rounds. Measured in candidate
evaluations the genetic arm uses a median of 720 per behavior and the second family 24, a
thirtyfold difference, 1.73 million evaluations against 57.6 thousand. Where the second
family performs \emph{worse}, budget is therefore the first explanation and we do not
claim a
mechanism; where it performs \emph{better}, budget cannot be the explanation.

\paragraph{The behavior-level verdict replicates.}
Within the second family the attack reproduces itself against the undefended model at
$\phi_{\mathrm{null}} = 0.296$, with the three seed-pair estimates 0.402, 0.309 and 0.178
spanning 0.223. Among the fifteen non-guard pairs $\phi$ runs from 0.033 to 0.326, one sits
above the null, and the largest excess is $+0.030$. That is the same verdict
Section~\ref{sec:e1} reaches on the genetic arm, where the largest excess is $+0.026$ against
a null of 0.620: the excess is a small fraction of the null's
own scatter. Conditioning on behavior difficulty removes 48\% of the mean association here
against 71\% there, and the random-intercept fit gives an intraclass correlation of 0.391
against 0.693. Two searches that share only their objective therefore agree that no pair of these
safeguards is distinguishable from the attack finding the same behaviors twice, which
indicates that the result concerns the stack and the
behavior set and not one optimizer.

\paragraph{The guard model's result does not replicate.}
Table~\ref{tab:family} gives attack success per member under both families. Six of the seven
are lower under the second family, by 0.46 to 0.68, which is what a thirtyfold budget
difference predicts.

The guard model runs the other way. Against the genetic adversary it
admits 0.030 of attacked prompts. Against the second family it admits \textbf{0.193}:
0.22, 0.17 and 0.19 across the three replicates, with Wilson intervals of $[0.016, 0.056]$
and $[0.150, 0.238]$ that do not overlap. Of the behaviors either family breached through the
guard, 55 were breached by the second family alone and 6 by the genetic arm alone, a pooled
exact McNemar $p = 5.4 \times 10^{-11}$.

This cannot be budget: the family that does better here is the one with a thirtieth of the
evaluations. The mechanism is visible in the prompts. The genetic arm builds its prompts by
recombining framing sentences, and a semantic guard classifies the result as a jailbreak
attempt; the attacker model rewrites the request itself into something the guard reads as
benign, and needs few attempts to find one.

\paragraph{Consequences.}
The guard's 0.03 is therefore specific to the adversary it was measured against. The guard
is not useless: under the second family 0.807 of attempts still fail against it. But the
member that a single-family evaluation identifies as the one worth keeping is
the member most sensitive to which family was used.

\paragraph{The families break different behaviors.}
Per member, agreement between the families on \emph{which} behaviors break is low: $\phi$
from 0.114 to 0.213 among the non-guard members, 0.066 for the guard. We read this cautiously,
because $\phi$ between two indicators with marginals of 0.22 and 0.74 is attenuated by the gap
in base rates alone. The asymmetry is interpretable. For
every member except the guard the behaviors breached by the genetic arm alone outnumber those
breached by the second family alone by an order of magnitude, which is the budget again; for
the guard the ratio inverts, 55 against 6.

\begin{table}[!htbp]
\centering
\caption{Attack success per member under the two attack families, pooled over three replicates of 100 behaviors, and the number of attempts breached through that member by both families or by one alone. The budgets are not matched: a median of 720 candidate evaluations per behavior for the genetic search and 24 for the LLM-driven one.}
\label{tab:family}
\small
\begin{tabular*}{\textwidth}{@{\extracolsep{\fill}}lccccc@{}}
\toprule
& \multicolumn{2}{c}{Attack success} & \multicolumn{3}{c}{Attempts breached by} \\
\cmidrule(lr){2-3}\cmidrule(lr){4-6}
Safeguard & Genetic & LLM-driven & Both & Genetic only & LLM only \\
\midrule
Refusal-priming prompt     & 0.75 & 0.07 & 20 & 205 & 2 \\
Probe, early layer         & 0.75 & 0.21 & 53 & 172 & 9 \\
Token-anomaly statistic    & 0.74 & 0.24 & 65 & 158 & 8 \\
Probe, middle layer        & 0.73 & 0.21 & 54 & 166 & 8 \\
Windowed likelihood filter & 0.72 & 0.24 & 64 & 152 & 8 \\
Perturbation aggregator    & 0.65 & 0.19 & 43 & 151 & 13 \\
\textbf{Guard model (8B)}  & \textbf{0.03} & \textbf{0.19} & 3 & 6 & \textbf{55} \\
\addlinespace
\emph{Undefended}          & \emph{0.74} & \emph{0.22} & 61 & 161 & 6 \\
\bottomrule
\end{tabular*}
\end{table}

\subsection{What the hidden-state probes detect}\label{sec:probes}

The two probes form the within-family pair of Section~\ref{sec:e1}, and their block rates
change most between attack classes and targets. We therefore ask directly what they
separate (Table~\ref{tab:probes}), instead of
reporting a validation
figure from a held-out split of their own training pool.

\begin{table}[!htbp]
\centering
\caption{What the mid-layer probe (probe$_{16}$) and the early-layer probe (probe$_{8}$) separate. Both corpora in the second row are harmless, so an AUROC above 0.5 there is a corpus effect. Block rates are realized at each probe's deployed threshold, on 100 prompts for the benchmark rows, 300 for the attacked row and 400 for the corpus rows. The score shift is the paired per-behavior mean.}
\label{tab:probes}
\small
\begin{tabular*}{\textwidth}{@{\extracolsep{\fill}}lcc@{}}
\toprule
Quantity & probe$_{16}$ & probe$_{8}$ \\
\midrule
AUROC, harmful vs.\ benign (same benchmark) & 0.933 & 0.643 \\
AUROC, two harmless corpora                 & 0.792 & 0.866 \\
\addlinespace
Blocked, plain harmful prompts              & 0.89 & 0.21 \\
Blocked, adaptively attacked prompts        & 0.00 & 0.00 \\
Score shift, plain to attacked              & $-66.4$ & $-24.6$ \\
\addlinespace
Blocked, calibration corpus                          & 0.018 & 0.013 \\
\quad the same threshold on benchmark benign prompts & 0.26 & 0.07 \\
\quad the same threshold on the over-refusal set     & 0.15 & 0.01 \\
\bottomrule
\end{tabular*}
\end{table}

Three things follow. The mid-layer probe is a harm detector on unmodified
prompts, separating harmful from benign at AUROC 0.933 and blocking 89 of 100
plain harmful prompts at its deployed threshold, and the adaptive attack
evades it completely: attacked prompts score 66.4 points lower than
plain harmful ones, and the entire attacked distribution falls below the
threshold, its maximum included, at $-10.5$ against a threshold of $-5.2$, so the
probe blocks none of the 300. The
early-layer probe separates two \emph{harmless} corpora better than it separates
harm (0.866 against 0.643), so it is substantially a topic detector. And an
operating point calibrated to block 1\% of a general instruction corpus blocks
26\% of the benchmark's own benign prompts, so the threshold does not transfer
across corpora.

\paragraph{The mid-layer probe on the aligned target.}
On Llama-2 the
mid-layer probe blocks 48 of 100 attacked prompts, against none of 100
on the primary target. The probe never blocks a prompt the search has found
to evade it, on either target: 0 of the 100 prompts the search produced on the primary
target, and
0 of the 52 it produced on Llama-2. The 48 are the behaviors for which the
search found \emph{no} evading prompt within budget, and they are not behaviors
the attack was otherwise winning, because the undefended run breached only 1 of
those 48.

What differs between the targets is the adversary, not the detector. The attack's
objective, the log-probability of the affirmative prefix, reaches a median of
$-0.001$ against the weakly aligned model and $-1.339$ against the aligned one:
saturated in the first case, far from it in the second. We therefore report the 48 as cells
in which no evading prompt was found within the stated search budget, and not as
the probe succeeding.

This bears on the within-family pair. The two probes have the largest $\phi$ in
Table~\ref{tab:e1}, 0.646, but one of them is substantially a corpus
detector and neither contributes a single block under attack on the primary target, so
whatever
the pair shares is not a view of harm that the attack has to defeat. We keep the pair in
the analysis, because removing a member after seeing the result
would be worse, but it should carry less weight than its position in
Table~\ref{tab:e1} suggests.

\section{Discussion}\label{sec:discussion}

The results bear on two questions that a practitioner would ask of a diversity estimate:
what it can be used for, and what a reported figure describes. We take them in turn and
then state the limits of the study.

\subsection{What a diversity measurement is for}

The ensemble literature treats diversity as a property to be engineered and then verified
\citep{brown2005diversity}, and the natural reading of that advice for a safeguard stack is
that measuring diversity between candidate members tells a designer which members to combine.
Our results limit that use. Pairwise $\phi$
among the non-guard members spans 0.488 to 0.646, which looks like a ranking of
which mechanisms share a blind spot. But no value in it is distinguishable from the correlation
between two independent runs of the same attack against the undefended model, and
conditioning on how hard
each behavior is to elicit removes 71\% of it, or 83\% by the fitted model, and only
partly preserves
the ordering. What is left is small: partial $\phi$ between 0.002 and
0.268, with intervals that overlap. At this sample size and
against this adversary, the measurement does not tell a designer which members are
near-substitutes.

The second stage does not depend on any of this. A stack whose components were each
evaluated separately has no measured robustness, because the adversary optimizes against the
composition, and the composition estimate erred in the same direction on both axes. On
security, the six-layer stack breaches 0.573 of behaviors where the
intersection of its own members predicts 0.473, with a pooled exact McNemar $p = 0.0018$ over
three replicates, and every one of the six stacks we attacked shows a gap in that direction
(Section~\ref{sec:e3}). On benign traffic, the seven-member stack refuses 0.830 of matched
benign prompts where the union of its members' own refusals is 0.600, identically in all three
replicates, and 0.995 of an independent benign sample (Section~\ref{sec:frr}). A per-layer
composition
estimate is therefore a best case, not a projection, and the gap comes from members that
interact, which measuring them one at a time cannot see. Per-layer measurements remain
worth taking, because
they are cheap and they show which members do anything at all
(Section~\ref{sec:e2}), but they should not be composed into an estimate for
the stack.

\subsection{Diversity is a property of a (stack, adversary, target) triple}

A defense's measured strength depends on the attack class it faces: the
same likelihood filter is a complete defense against one adversary and inert against
another, and the guard model admits 0.03 under one attack family and 0.19 under the other.
Every diversity estimate containing such a member inherits the dependence, so a diversity
figure for a
stack is uninterpretable without the adversary that produced it, and composition estimates
assembled from per-defense figures obtained under different attacks can be optimistic without
any bound that we can state.

The replication adds the target. On identical inputs the
perturbation aggregator lowers attack success on the weakly aligned target and does not on
the aligned one, and the mid-layer probe's block rate on fluent prompts goes from none to
about half (Table~\ref{tab:attackclass}; Section~\ref{sec:replication}). The
unit of description is therefore a triple of the stack, the adversary and the target, and a
figure quoted without all three does not transfer.

The dependability analogy stops here. Coincident-failure
modeling for redundant components \citep{littlewood2001modeling, littlewood2004redundancy}
assumes a fixed, if unknown, distribution of demands. Here the demand distribution is chosen
by an opponent who can see what the members share, so the quantity being estimated moves when
the opponent does. The appropriate response is to report
the adversary alongside the estimate, and to treat an estimate from a weaker adversary as an
upper bound
on diversity.

\subsection{Limitations}

The following points limit what the results establish.

\paragraph{Cause of the shared failures.}
We expected a shared substrate to be the explanation: every member wraps the same
target network, and safety behavior in such a network is shallow \citep{qi2024safety}. The
measurement does not require it. Most of the association (71\%) is accounted for by how hard
each behavior is to elicit from the undefended model, and what remains does
not exceed the reproducibility of the attack. The competing explanations are left standing.
Benchmark
construction is one, since the harmful and benign prompts are built as matched counterparts and
may share structure no mechanism would. The shared autograder is another, which
Section~\ref{sec:asr} bounds but does not eliminate. Establishing or refuting the substrate
claim requires varying the substrate while holding the defenses fixed, which this
design does not do. The claim implies a testable prediction: if a shared
substrate is responsible for the residue, deepening the target's alignment should lower the
correlation between members that are otherwise unchanged.

\paragraph{Informative regime.}
Correlation between two binary indicators is undefined when either is constant, so a safeguard
the available adversary cannot stress cannot be assessed for correlated failure. This already
happens on the weakly aligned
target, where the guard model blocks 96\% of the attacked prompts, its breach vector is nearly
constant, and every pair containing it collapses toward zero, and it happens more widely on
the aligned target (Section~\ref{sec:replication}). This
family of analysis is therefore informative where the defenses are
weak and uninformative where they are strong. A measurement regime that stays informative
against effective safeguards is an open problem that we do not solve.

\paragraph{A prediction recorded before the replication.}
Before the second target was attacked we recorded the prediction that on a refusal-robust
model the marginal breach
rates would approach zero and most pairs would become undefined. It was half right. The
guard model became degenerate as expected. The other six members stayed measurable, with
marginal attack success between 0.05 and 0.50 and test-retest reliabilities
between 0.31 and 0.47, and all twenty-one pairs are defined. The degeneracy is therefore a
property of
the individual member's operating point, not of the target.

\paragraph{Scope.}
The study covers single-turn jailbreaks. It uses two open-weight
targets of the same scale, one behavior benchmark and two attack families, each a single
optimizer and not an
opponent free to choose among strategies. Nothing here
establishes that the same structure holds at frontier scale, on proprietary stacks, in
multi-turn settings, or for prompt injection as opposed to jailbreak.

\paragraph{Grading.}
Breach labels come from an automated rubric grader whose variability we measure, with human
adjudication on a targeted subset. That
adjudication has a single annotator, who is an author, so it compares the grader with one
informed reader and not with a consensus of readers; the test-retest figure bounds that
reader's own consistency only. The
agreement figures in Section~\ref{sec:asr} bound how much of the reported structure could
be grader artifact; they do not eliminate the possibility.

\paragraph{Specification and deviations.}
The hypotheses, the dependency families and the primary pairs were fixed before the first run
and recorded in a preregistration document, which is included in the repository with its
dated amendment. We depart from it in two respects.
One pair was added during the work, after the originally designated within-family pair proved
unmeasurable because one of its members admitted nothing under the first adversary. And the
refusal hypothesis as registered concerned whether the stack's false-refusal rate tracks the
independence prediction; what we report is the broader claim that it is the binding constraint
on depth, which is a different statement and is labeled as such.

\section{Conclusion}\label{sec:conclusion}

Stacking safeguards around a language model is an ensemble argument, and it requires
members that fail on different
inputs. That condition is three quantities: separately optimized attacks support a
claim about which behaviors are jointly vulnerable, the identity governing a deployed
stack is written over shared inputs, and neither predicts what an adversary achieves
against the assembled pipeline. We defined the three estimands, gave each an
estimator and a control, and measured all three on a seven-member stack.

At the behavior level, two runs of the same attack
against the undefended model agree at $\phi_{\mathrm{null}} = 0.620$, and no safeguard
pair exceeds that value by more than the scatter among the null's own three estimates.
Behavior difficulty accounts for 71\% of the pairwise association on the partial
correlation and 83\% on the random-intercept model. The members are not independent,
but this design
cannot attribute their dependence to the pairing of mechanisms. On shared inputs, four of
the seven
members return the target's own response to nearly every adversarial prompt, which
only a cross-evaluation that also scores the undefended arm reveals. Attacked as
assembled, the six-layer stack breaches more often than the intersection of its members
predicts, 0.573 against 0.473 (pooled exact McNemar $p = 0.0018$), and the assembled
seven-member stack refuses 0.83 of matched benign prompts where the union of its
members' refusals is 0.60.

A member's measured strength belongs to a (stack, adversary, target) triple. Under the
genetic attack the guard model reduces attack success from 0.74 to 0.03; under a
second attack family with one thirtieth of the budget it admits 0.19. On identical inputs
the perturbation aggregator lowers attack success on one target and not on the other.
The analysis is also informative only where safeguards sometimes fail: on
the more strongly aligned target the attack succeeds on too few
behaviors for the null to be estimated precisely, and a measurement that stays informative
against effective safeguards remains an open problem.

Three recommendations follow for practice. Report the adversary alongside any diversity
figure. Report the null, because a pair statistic without it cannot be
distinguished from the attack finding the same behaviors twice. And measure the
assembled configuration, because a stack whose members were each evaluated alone has no
measured robustness.

\backmatter

\section*{Statements and Declarations}

\bmhead{Funding}
This work was supported by the Saudi Data and AI Authority (SDAIA) and King Fahd
University of Petroleum \& Minerals (KFUPM) under the SDAIA-KFUPM Joint Research
Center for Artificial Intelligence Grant JRC-AI-RFP-20, and by grant CRPG-25-2057
under the Cybersecurity Research and Innovation Pioneers Initiative of the
National Cybersecurity Authority (NCA) in the Kingdom of Saudi Arabia.

\bmhead{Competing interests}
The authors declare that they have no known competing financial interests or
personal relationships that could have appeared to influence the work reported
in this paper.

\bmhead{Ethics approval and consent to participate}
Not applicable. The study involves no human participants or personal data.

\bmhead{Data availability}
All derived data are available at
\url{https://github.com/AbrarAlotaibi/defense-correlation}. The repository contains the
per-behavior breach vector behind every estimate in the paper and the grading rubric
with the digest used to key the verdict cache. The adversarial prompts the search
produced are included, because the measurements cannot be reproduced or audited
without them.
The target models, the behavior benchmark and the external calibration corpus are
third-party artifacts, cited where they are introduced and not redistributed here.

\bmhead{Code availability}
All code is in the same repository: the safeguard implementations with their
calibrated thresholds, both attack harnesses, and the analysis scripts that regenerate
every table and figure from the stored verdicts.

\bmhead{Author contribution}
\textbf{Abrar Alotaibi:} Conceptualization, Methodology, Software,
Investigation, Formal analysis, Data curation, Validation, Visualization,
Writing -- original draft, Writing -- review \& editing.
\textbf{Moataz Ahmed:} Supervision, Funding acquisition, Project administration,
Writing -- review \& editing.

\begin{appendices}

\section{Per-replicate bootstrap intervals}\label{app:bootstrap}

Section~\ref{sec:e1} supports the claim that no pair is distinguishable from the null
with the median width of the bootstrap intervals on the pair
estimates. Table~\ref{tab:bootstrap_v2_vicuna} reports the intervals themselves. No
non-guard interval lies entirely above the mean null of 0.620, and three of the 45 lie
entirely below it.

\begin{table}[!htbp]
\centering
\caption{Behavior-level $\phi$ for every pair in each replicate, with bootstrap 95\% intervals from 2000 resamples over behaviors. The median width of these 63 intervals is 0.343.}
\label{tab:bootstrap_v2_vicuna}
\footnotesize
\begin{tabular*}{\textwidth}{@{\extracolsep{\fill}}lccc@{}}
\toprule
Pair & Replicate 1 & Replicate 2 & Replicate 3 \\
\midrule
probe$_{16}$ $|$ probe$_{8}$ & 0.637 [0.46, 0.81] & 0.583 [0.38, 0.75] & 0.718 [0.55, 0.87] \\
probe$_{8}$ $|$ priming & 0.682 [0.51, 0.84] & 0.566 [0.36, 0.74] & 0.646 [0.47, 0.81] \\
probe$_{16}$ $|$ anomaly & 0.585 [0.40, 0.76] & 0.617 [0.43, 0.78] & 0.689 [0.51, 0.84] \\
probe$_{16}$ $|$ priming & 0.666 [0.49, 0.83] & 0.646 [0.46, 0.81] & 0.574 [0.39, 0.75] \\
filter $|$ priming & 0.699 [0.55, 0.84] & 0.615 [0.42, 0.78] & 0.566 [0.38, 0.74] \\
probe$_{8}$ $|$ anomaly & 0.645 [0.46, 0.81] & 0.474 [0.26, 0.66] & 0.711 [0.53, 0.86] \\
filter $|$ anomaly & 0.549 [0.37, 0.72] & 0.689 [0.51, 0.85] & 0.573 [0.38, 0.75] \\
priming $|$ anomaly & 0.682 [0.51, 0.84] & 0.553 [0.36, 0.73] & 0.566 [0.39, 0.74] \\
filter $|$ probe$_{16}$ & 0.545 [0.37, 0.72] & 0.674 [0.49, 0.82] & 0.533 [0.34, 0.72] \\
filter $|$ probe$_{8}$ & 0.645 [0.46, 0.80] & 0.547 [0.34, 0.72] & 0.553 [0.36, 0.73] \\
probe$_{8}$ $|$ aggregator & 0.622 [0.47, 0.76] & 0.515 [0.32, 0.68] & 0.521 [0.34, 0.70] \\
filter $|$ aggregator & 0.550 [0.38, 0.71] & 0.693 [0.53, 0.83] & 0.351 [0.16, 0.54] \\
priming $|$ aggregator & 0.494 [0.33, 0.65] & 0.620 [0.44, 0.78] & 0.476 [0.29, 0.65] \\
aggregator $|$ anomaly & 0.437 [0.25, 0.61] & 0.594 [0.42, 0.75] & 0.545 [0.37, 0.71] \\
probe$_{16}$ $|$ aggregator & 0.533 [0.37, 0.69] & 0.527 [0.34, 0.70] & 0.404 [0.21, 0.58] \\
\addlinespace
guard $|$ filter & 0.121 [0.06, 0.19] & 0.107 [0.05, 0.17] & 0.102 [0.05, 0.16] \\
guard $|$ probe$_{16}$ & 0.102 [0.05, 0.16] & -0.021 [-0.24, 0.13] & 0.107 [0.05, 0.17] \\
guard $|$ priming & 0.091 [0.05, 0.15] & 0.104 [0.05, 0.16] & -0.021 [-0.23, 0.14] \\
guard $|$ probe$_{8}$ & -0.025 [-0.24, 0.13] & 0.093 [0.05, 0.15] & 0.104 [0.05, 0.17] \\
guard $|$ aggregator & 0.020 [-0.18, 0.17] & 0.121 [0.06, 0.19] & 0.006 [-0.19, 0.16] \\
guard $|$ anomaly & -0.025 [-0.25, 0.13] & -0.034 [-0.25, 0.12] & 0.102 [0.05, 0.16] \\
\bottomrule
\end{tabular*}
\end{table}

\end{appendices}

\bibliography{refs}

\end{document}